%% file: main.tex
\makeatletter\let\kernel@label\label\makeatother
\documentclass[aps,prb,reprint,nofootinbib,longbibliography,floatfix]{revtex4-2}

\usepackage[utf8]{inputenc}
\usepackage[T1]{fontenc}
\usepackage{lmodern}
\usepackage{amsmath,amssymb,amsthm,mathtools}
\usepackage{booktabs}
\usepackage{algorithm,algorithmic}
\makeatletter
\def\ftype@algorithm{8}
\let\endalgorithm\float@end@ltx
\let\float@end\float@end@ltx

\let\label\kernel@label
\def\@bibdataout@rev{%
 \immediate\write\@bibdataout{@CONTROL{REVTEX42Control\eprint@enable@sw{,eprint="0"}{,eprint="1"}}}%
 \if@filesw\immediate\write\@auxout{\string\citation{REVTEX42Control}}\fi}
\makeatother
\usepackage{enumitem}
\usepackage{xcolor}
\usepackage{graphicx}
\usepackage{tikz}
\usepackage{pgfplots}
\pgfplotsset{compat=1.18}
\input{figures/tikz/styles.tex}
\usepackage{adjustbox}
\usepackage{microtype}
\usepackage{xurl}
\usepackage[colorlinks=true,linkcolor=blue!50!black,citecolor=blue!50!black,urlcolor=blue!50!black]{hyperref}
\usepackage{orcidlink}
\usepackage[capitalise]{cleveref}
\crefname{section}{Sec.}{Secs.}
\crefname{subsection}{Sec.}{Secs.}
\Crefname{section}{Section}{Sections}
\Crefname{subsection}{Section}{Sections}
\crefname{figure}{Fig.}{Figs.}
\Crefname{figure}{Figure}{Figures}
\crefname{algorithm}{Algorithm}{Algorithms}

\makeatletter
\def\p@subsection{}
\def\p@subsubsection{}
\makeatother
\newtheorem{proposition}{Proposition}
\newcommand{\C}{\mathbb{C}}
\newcommand{\ord}{\mathcal{O}}
\newcommand{\code}[1]{\texttt{#1}}
\newcommand{\cc}[1]{\overline{#1}}
\newcommand{\GB}{\,\mathrm{GB}}
\newcommand{\s}{\,\mathrm{s}}
\newcommand{\TFs}{\,\mathrm{TFLOP/s}}
\newcommand{\orth}{\operatorname{orth}}

\makeatletter
\@booleantrue\raggedcolumn@sw

\makeatother
\begin{document}

\title{\texttt{portc}: Parallel Out-of-Core Randomized Tensor-Network Compression}
\author{Ram\'on L. Panad\'es-Barrueta\,\orcidlink{0000-0003-4239-0978}}
\author{Davide Materia\,\orcidlink{0009-0008-5082-8902}}
\author{Matteo A. C. Rossi\,\orcidlink{0000-0003-4665-9284}}
\author{Roberto Di Remigio Eik{\aa}s\,\orcidlink{0000-0002-5452-9239}}
\affiliation{Algorithmiq S.r.l., Via della Chiusa 15, 20123 Milan, Italy}

\begin{abstract}
  We present \code{portc}, an open-source multi-GPU and out-of-core implementation of
  successive randomized compression (SRC)~\cite{camano2026src} that extends randomized
  tensor-train rounding~\cite{aldaas2023randomized,aldaas2026khatrirao} to \emph{stacks}
  of several tensor trains without forming their full product. Partitioning the
  structured random projections distributes contractions and orthogonalization across
  GPUs while staging intermediates across device memory, host memory and disk, supported
  by an analytical cost model and an exact compression-error identity. On a 50-site
  stack compressed from bond dimension $256\,000$ to $2000$, \code{portc} achieves
  speed-ups over a single device of $3.8\times$ on four NVLink-connected GH200s and
  $5.8\times$ on eight PCIe-connected RTX PRO 6000 cards. A Heisenberg-picture simulation of a
  127-qubit kicked Ising circuit reaches operator bond dimension $8192$, to our
  knowledge the largest reported for a matrix-product simulation of this circuit,
  illustrating both norm-loss diagnostics and the limits of a one-dimensional ansatz.
  Overall, the structure of SRC
  enables large tensor-train contractions to be distributed across devices and memory
  hierarchies with quantitatively predictable performance, while the experiments
  clarify when stacked compression improves the accuracy--cost trade-off over pairwise
  compression.
\end{abstract}

\maketitle

\section{Introduction}\label{sec:intro}

A recurring operation in tensor-network simulations is to apply an operator and
compress the result before its representation becomes too large. For a matrix
product state (MPS), this operation advances a time evolution, applies circuit
layers or evaluates operator actions~\cite{paeckel2019timeevolution}; for a matrix
product operator (MPO), it composes propagators, evolves observables or builds quantum
channels~\cite{schollwock2011dmrg}. The tensor-train (TT)
format~\cite{oseledets2011tt} avoids storing exponentially large states and
operators, but not large intermediates: the product of trains with bond dimensions
$D_1,\dots,D_k$ has bond dimension $\prod_{i}D_i$ before compression.

This bottleneck affects both quantum many-body simulations and the classical
processing that accompanies quantum experiments. Validating quantum dynamics can
require tensor-network calculations at the largest affordable bond dimension,
including multi-GPU simulations~\cite{barron2026observable}. Tensor-network error
mitigation (TEM) poses a related workload, trading quantum sampling overhead for
classical tensor-network
post-processing~\cite{filippov2023tem,mangini2024noise,filippov2024scalability}. In
both settings, efficient compression makes
larger bond dimensions accessible but does not by itself guarantee convergence of the
physical prediction.

The conventional approach forms the product and rounds it by sequential
singular value decompositions (SVDs), at a bulk cost cubic in the uncompressed bond
dimension for fixed physical dimension~\cite{oseledets2011tt}. Zip-up contraction~\cite{stoudenmire2010zipup},
 randomized tensor-train
rounding~\cite{aldaas2023randomized,kressner2023streaming,aldaas2026khatrirao} and
successive randomized compression (SRC)~\cite{camano2026src} avoid some or all of
this intermediate growth. Without forming the product, SRC contracts the open
physical legs with random test tensors (a sketch) in a left-to-right pass to build
sketched prefix environments, and extracts the orthonormal cores (output bases) of
the compressed train in a right-to-left pass. For an MPO--MPS product, its range-finding cost is quadratic in the target
bond dimension and linear in the product of the input bond dimensions. SRC has
already been used to compress the Heisenberg-picture product $U^\dagger O U$ in
simulations of a 120-qubit Fermi--Hubbard experiment~\cite{ouyang2026hubbard}.
Complementary work parallelizes variational MPO multiplication by distributing
segments of the chain~\cite{fodera2026parallel}.

To support large-scale tensor-network simulations, we introduce \code{portc},
an open-source implementation of SRC for CPU and GPU architectures.
Although SRC avoids forming the product tensor, it stores $n-1$
prefix environments across the chain, which at large bond dimensions can exceed
single-GPU and host memory. The random sketch index provides a natural dimension for
distribution across GPUs. Together with the output bond index derived from it, it is
a free index in every contraction, so the contractions and the stored environments can
be partitioned across devices and batched out of core. Only orthogonalization couples
the sketch columns; the other collectives replicate the projected environment and the
input cores. Exploiting this
structure at scale requires care with orthogonalization, communication and memory
management. The main contributions of this work are:
\begin{enumerate}[leftmargin=*]
  \item \textbf{SRC for stacks.} We formulate the sweep for a stack of $k$ trains
        (MPOs, optionally terminated by an MPS), with optional oversampling followed
        by a deterministic SVD truncation.
        We prove a deterministic error identity for the sweep and analyse when
        compressing a stack in one sweep pays off over pairwise application.
  \item \textbf{A distributed sweep.} The sketch index is uncontracted in every
        contraction except the QR\@. A row-distributed shifted Cholesky QR exchanges
        only Gram matrices when the sketch is sufficiently well conditioned, with a
        Householder QR as fallback. The environment contractions of the first pass
        need no communication, the stored environments are partitioned across devices,
        and the random sketches are independent of the device count.
  \item \textbf{Out-of-core execution.} A planner sizes every contraction to the
        device budget by batching along the free index and assigns each
        environment to device memory, host memory or disk; a staged pipeline overlaps
        the transfers with computation. For fixed sketches, batching and data
        placement leave the result unchanged in exact arithmetic.
  \item \textbf{Cost model and experiments.} We model the arithmetic, memory and
        communication costs of each step and benchmark \code{portc} across three
        multi-GPU platforms: NVLink-connected GH200 nodes on JUPITER and A100 nodes on
        Leonardo, and an eight-GPU PCIe node. From a single-GPU run and link
        microbenchmarks alone, the parameter-free model predicts four-GPU run times on
        JUPITER within $2\%$. The experiments reach an out-of-core stack whose
        working set exceeds host memory and a 127-qubit circuit evolved at bond
        dimension 8192, clarifying when single-sweep stacking outperforms pairwise
        compression.
\end{enumerate}

\Cref{sec:stacks} introduces the notation and randomized range finding, formulates
SRC for stacks and counts its operations. \Cref{sec:parallel,sec:ooc} describe the
distributed sweep and its out-of-core execution, and \cref{sec:model} combines their
costs into a performance model. \Cref{sec:experiments} evaluates the implementation,
\cref{sec:application} applies it to a circuit implementing a kicked Ising model, and
\cref{sec:conclusions} discusses limitations and future directions. The appendices
summarize the notation, path diagnostics, supplementary benchmarks, Cholesky QR
pseudocode and memory footprint.

\section{Successive randomized compression of stacks}\label{sec:stacks}

\begin{figure*}[t]
  \centering
  \begin{adjustbox}{max width=\linewidth}\input{figures/tikz/src_steps.tex}\end{adjustbox}
  \caption{SRC on the $n$-site MPO--MPS stack $A^{(1)}A^{(2)}\psi$ (green: MPO cores;
    blue: MPS core; yellow: Gaussian sketch; the unit lower leg $x_k$ of $\psi$ is
    omitted).\ (a) Full stack across sites $1,\dots,n$, highlighting bulk site $j$.\
    (b) Environment update~\eqref{eq:ltr}; the dot joins the sketch
    indices $s$ of the prefix environments $C_j$, $C_{j+1}$ and $\Omega_j$.\ (c)
    Sketch~\eqref{eq:sketch} combining $C_j$, the stack cores and the suffix
    environment $S_{j+1}$, with open legs $s$, $x_0$ and $\eta$.\ (d) Thin QR
    factorization of the reshaped sketch $M_j$, with rows $(\eta,x_0)$ and columns
    $s$, yielding the right-isometric output core $\Gamma_j$.\ (e)
    Projection~\eqref{eq:project} forming $S_j$ (overbars denote complex
    conjugation).}\label{fig:src-steps}
\end{figure*}
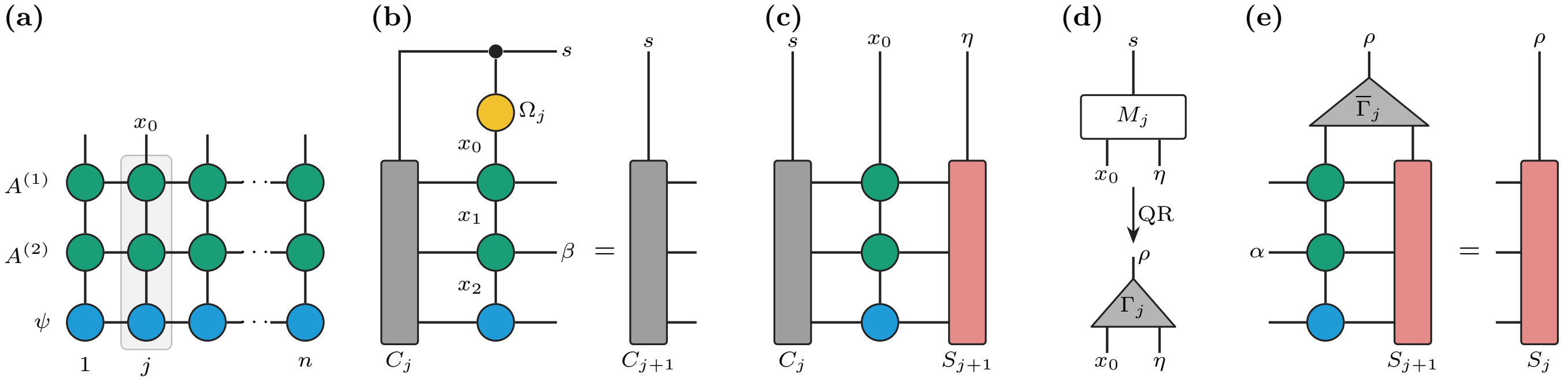

SRC builds the compressed train by a sequence of randomized range-finding steps.
Given a matrix $T$, the randomized range finder~\cite{halko2011finding} draws a
Gaussian test matrix $\Omega$ with $\ell$ columns, forms an orthonormal basis
$Q=\orth(T\Omega)$ for the column space of $T\Omega$, and approximates
$T\approx QQ^\dagger T$. Choosing $\ell$ larger than the target rank $\chi$
(oversampling) improves accuracy when the singular values decay, with probabilistic
error bounds controlled by the tail beyond rank
$\chi$~\cite{halko2011finding,martinsson2020rnla}. SRC applies this idea to successive
unfoldings of the product using a structured test matrix that is never formed
explicitly. A sweep consists of two passes (\cref{fig:src-steps}), plus a third when
the sketch is oversampled.

\subsection{Stacks of tensor trains}

A \emph{stack} is a product of $k$ tensor trains on $n$ sites in ket form: MPOs,
optionally terminated by an MPS\@. Layer $i\in\{1,\dots,k\}$ at site
$j\in\{1,\dots,n\}$ is a four-index core
\begin{equation}
  A^{(i)}_j[a_i, b_i, x_{i-1}, x_i],
\end{equation}
with left bond $a_i$, right bond $b_i$, upper physical leg $x_{i-1}$ and lower
physical leg $x_i$, so that consecutive layers share a physical leg. Boundary bonds
and the lower leg of an MPS layer have dimension one, giving every core in the stack
the same four-index form. The open legs of the product are $x_0$
(upper) and $x_k$ (lower); if the last layer is an MPS, only $x_0$ remains. A stack
that begins with a row vector is handled by transposing the product and reversing the
layers.

The multi-indices $\alpha=(a_1,\dots,a_k)$ and $\beta=(b_1,\dots,b_k)$ group the
left and right bonds across layers at site $j$, with uncompressed product bond
dimensions $\Pi_{D,j}=\prod_i\dim a_i$ on the left and $\Pi_{D,j+1}=\prod_i\dim b_i$
on the right ($\Pi_{D,1}=\Pi_{D,n+1}=1$). The open physical dimension of site $j$ is
$p_j=\dim x_0\dim x_k$. For uniform bulk cost estimates we write $D_i$ for the bulk
bond dimension of layer $i$, $\Pi_D=\prod_{i}D_i$ for their product,
$\Sigma_D=\sum_{i}D_i$ for their sum, $d$ for each physical leg dimension, and $p=d^2$
for an MPO output ($p=d$ for an MPS). \Cref{app:notation} summarizes the notation.

\subsection{Left to right: sketched environments}

The first pass sweeps left to right to build sketched prefix environments
$C_j\in\C^{\ell\times \Pi_{D,j}}$ ($j=1,\dots,n$), where $C_j$ is the exact product of
sites $1,\dots,j-1$ with all open physical legs contracted against random test
tensors. Draw independent tensors $\Omega_j\in\C^{\ell\times\dim x_0\times\dim x_k}$,
$j=1,\dots,n-1$, with standard complex Gaussian entries (real ones for real inputs),
initialize the trivial left boundary $C_1\in\C^{\ell\times 1}$ by $C_1[s,1]=1$, and
for $j=1,\dots,n-1$ compute
\begin{equation}\label{eq:ltr}
  \begin{split}
    C_{j+1}[s, \beta] = \sum_{\alpha,\,x_0,\dots,x_k}
    &C_j[s, \alpha]\; \Omega_j[s, x_0, x_k]\\
    &\times\prod_{i=1}^{k} A^{(i)}_j[a_i, b_i, x_{i-1}, x_i].
  \end{split}
\end{equation}
Because all sites share the sketch index $s\in\{1,\dots,\ell\}$, the implied test
matrix is a columnwise Kronecker (Khatri--Rao) product of the
$\Omega_j$~\cite{camano2026src,aldaas2026khatrirao}. Its columns are independent,
whereas the entries within each column are coupled, so standard Gaussian error bounds
do not apply directly~\cite{camano2026src}. Although worst-case subspace bounds for
Khatri--Rao sketches degrade exponentially with the chain length, their empirical
accuracy is close to that of dense Gaussian sketches~\cite{camano2025osi}.

\subsection{Right to left: range finding and projection}\label{sec:rtl}

The second pass sweeps right to left to extract right-isometric output cores
$\Gamma_j$ of left bond dimension $r_j\le\ell$ and right bond dimension
$\eta_j=r_{j+1}$, together with projected suffix environments
$S_j\in\C^{r_j\times \Pi_{D,j}}$ representing sites $j,\dots,n$ projected onto
$\Gamma_j,\dots,\Gamma_n$. Initialize the trivial right boundary $r_{n+1}=1$ and
$S_{n+1}=[1]\in\C^{1\times 1}$, and for $j=n,\dots,2$:
\begin{enumerate}[leftmargin=*]
  \item \textbf{Sketch.} For $\eta\in\{1,\dots,\eta_j\}$ and $s\in\{1,\dots,\ell\}$,
        compute
        \begin{equation}\label{eq:sketch}
          \begin{split}
            Y_j[\eta, x_0, x_k, s] = \sum_{\alpha,\beta,\,x_1,\dots,x_{k-1}}
            &C_j[s, \alpha]\; S_{j+1}[\eta, \beta]\\
            &\times\prod_{i=1}^{k} A^{(i)}_j[a_i, b_i, x_{i-1}, x_i],
          \end{split}
        \end{equation}
        and reshape it into the matrix $M_j\in\C^{(\eta_{j}p_j)\times\ell}$ with rows
        $(\eta,x_0,x_k)$.
  \item \textbf{Orthogonalize.} Compute a thin QR factorization $M_j=Q_{j}R_j$, where
        $Q_j$ has $r_j=\min(\eta_{j}p_j,\ell)$ columns. With a relative cutoff $\tau>0$,
        compute the SVD $R_j=U\Sigma V^\dagger$ and replace $Q_j$ by $Q_{j}U_{[:,1:r_j]}$,
        keeping the $r_j\ge 1$ singular values with $\sigma_i\ge\tau\sigma_1$ (note
        that $\sigma_i$ are singular values of the sketch $M_j$ rather than Schmidt
        values of the full product). Reshaping $Q_j$ gives the right-isometric core
        $\Gamma_j[\rho,\eta,x_0,x_k]=Q_j[(\eta,x_0,x_k),\rho]$ for
        $\rho\in\{1,\dots,r_j\}$.
  \item \textbf{Project.} Update the suffix environment $S_j\in\C^{r_j\times \Pi_{D,j}}$,
        where the overbar denotes complex conjugation:
        \begin{equation}\label{eq:project}
          \begin{split}
            S_j[\rho, \alpha] = \sum_{\eta,\beta,\,x_0,\dots,x_k}
            &\cc{\Gamma_j[\rho, \eta, x_0, x_k]}\; S_{j+1}[\eta, \beta]\\
            &\times\prod_{i=1}^{k} A^{(i)}_j[a_i, b_i, x_{i-1}, x_i].
          \end{split}
        \end{equation}
\end{enumerate}
Finally, the first core absorbs $S_2$,
\begin{equation}\label{eq:first}
  \begin{split}
    \Gamma_1[\rho, x_0, x_k] = \sum_{\beta,\,x_1,\dots,x_{k-1}}
    &S_2[\rho, \beta]\\
    &\times\prod_{i=1}^{k} A^{(i)}_1[1, b_i, x_{i-1}, x_i].
  \end{split}
\end{equation}
The output $\Gamma_1\Gamma_2\cdots\Gamma_n$ is right-canonical with bond dimensions
$r_j\le\ell$. For a stack of depth $k$, the four
contractions~\eqref{eq:ltr}--\eqref{eq:first} are fixed einsum expressions that cover
every site and every type of stack.

The sketch step is an instance of randomized range finding~\cite{halko2011finding}, although this 
connection is not immediately apparent. Indeed, let
$T_j$ be the product of sites $1,\dots,j$ contracted with $S_{j+1}$, unfolded into a
matrix whose rows are the indices $(\eta,x_0,x_k)$ of the new core and whose columns are
the combined open physical indices $\xi_l=(x_{0,l},x_{k,l})$ of all sites $l<j$. The
output core should span the dominant column space of $T_j$, but $T_j$ has exponentially
many columns and an SVD of it is out of reach. Instead, $M_j=T_j\Omega_{<j}$, where the
Khatri--Rao test matrix
$\Omega_{<j}[(\xi_1,\dots,\xi_{j-1}),s]=\prod_{l<j}\Omega_l[s,\xi_l]$ has only $\ell$
columns, so the QR factorization of $M_j$ returns an approximate basis of that column
space. 

\subsection{Oversampling and the truncation sweep}\label{sec:oversampling}

With $\ell=\chi$, the two passes return the output directly. Oversampling, $\ell>\chi$,
for instance $\ell=\max(\lceil1.5\chi\rceil,\chi+10)$ as in
Ref.~\cite{camano2026src}, captures the dominant subspace more accurately, and a
third, left-to-right pass truncates the intermediate to $\chi$. Because the
intermediate train is already right-canonical, no preliminary canonicalization sweep is
needed. At each site $j=1,\dots,n-1$, we reduce the core by a thin QR factorization,
compute the SVD of the small triangular factor to obtain the Schmidt values at bond
$j$, keep at most $\chi$ values above cutoff $\tau$ (applied here instead of in the
second pass), and absorb $\Sigma V^\dagger$ into site $j+1$. The output is
left-canonical, and the discarded weights add in quadrature~\cite{oseledets2011tt}. The
truncation pass costs $\ord(\ell^3 p)$ per site, subleading to the
$\ord(\ell^2p\Pi_D)$ range-finding term when $\Pi_D\gg\ell$, though both sketching
passes run at the larger size $\ell$.

\subsection{A deterministic error identity}\label{sec:error}

Let $\mathcal{T}$ be the exact product and $\widehat{\mathcal{T}}$ the output of the
two-pass sweep, with the open physical legs of each site grouped into a single index of
dimension $p_l$. Let $p_{<j}=\prod_{l<j}p_l$ and $p_{\ge j}=\prod_{l\ge j}p_l$.
Contracting the right-canonical cores $\Gamma_j,\dots,\Gamma_n$ defines an isometry
$U_j\in\C^{p_{\ge j}\times r_j}$ and the orthogonal projector
\begin{equation}
  \Pi_j = I_{p_{<j}} \otimes U_{j}U_j^\dagger, \qquad \Pi_{n+1} = I.
\end{equation}

\begin{proposition}\label{prop:error}
  In exact arithmetic, $\widehat{\mathcal{T}}=\Pi_2\mathcal{T}$ and
  \begin{equation}\label{eq:error}
    \begin{split}
      \|\mathcal{T}-\widehat{\mathcal{T}}\|_F^2
      &=\sum_{j=2}^n \|(\Pi_{j+1}-\Pi_j)\mathcal{T}\|_F^2\\
      &=\sum_{j=2}^n \|(I-Q_{j}Q_j^\dagger)T_j\|_F^2.
    \end{split}
  \end{equation}
\end{proposition}
\begin{proof}
  The relation $\widehat{\mathcal{T}}=\Pi_2\mathcal{T}$ follows
  from~\eqref{eq:project} and~\eqref{eq:first}. Each $U_j$ is $U_{j+1}$ extended by the
  isometry $\Gamma_j$, so the ranges of the $\Pi_j$ are nested,
  $\operatorname{range}\Pi_j\subseteq\operatorname{range}\Pi_{j+1}$. Differences of
  consecutive nested orthogonal projectors have mutually orthogonal ranges, and
  $\mathcal{T}-\Pi_2\mathcal{T}=\sum_{j=2}^n(\Pi_{j+1}-\Pi_j)\mathcal{T}$ telescopes,
  so the first equality in~\eqref{eq:error} follows from Pythagoras' theorem. For the
  second, $(\Pi_{j+1}-\Pi_j)\mathcal{T}$ is $(I-Q_{j}Q_j^\dagger)T_j$ mapped through
  the isometry $U_{j+1}$, which preserves the Frobenius norm.
\end{proof}

For any sketch, the squared error of the two-pass sweep equals the sum of the squared
local projection residuals in~\eqref{eq:error}. The identity does not by itself bound
these residuals a priori: the $T_j$ depend on suffix bases built from the same sketches.
Probabilistic bounds require an analysis of this dependence, which we do not attempt here (see
Ref.~\cite{camano2026src} for exact-recovery guarantees and leave-one-out error
estimators, and Ref.~\cite{aldaas2026khatrirao} for Khatri--Rao residual estimators).
Orthogonality also implies
\begin{equation}\label{eq:norm-loss}
  \|\mathcal{T}-\widehat{\mathcal{T}}\|_F^2
  =\|\mathcal{T}\|_F^2-\|\widehat{\mathcal{T}}\|_F^2,
\end{equation}
so the loss in squared norm gives the squared error of this single compression when
the norm of the uncompressed product is known. In unitary operator evolution, this is
the error of compressing the current iterate, not the accumulated error relative to
the exact evolution (\cref{sec:app-physics}). With oversampling, the triangle inequality gives
$\|\mathcal{T}-\widetilde{\mathcal{T}}\|_F\le
  \|\mathcal{T}-\widehat{\mathcal{T}}\|_F+\|\widehat{\mathcal{T}}-\widetilde{\mathcal{T}}\|_F$
for the truncated output $\widetilde{\mathcal{T}}$, where the second term is the square
root of the sum of squares of the Schmidt values discarded at the successive truncation steps.

\subsection{Operation counts}\label{sec:flops}

The contractions~\eqref{eq:ltr}--\eqref{eq:project} are evaluated pairwise, and their
cost depends on the contraction path. For fixed
physical dimension and one dominant input bond, absorbing that layer first gives the
leading bulk costs in complex multiply--adds,
\begin{subequations}\label{eq:flops}
  \begin{align}
    W_{\mathrm{env}}    & \approx\ell\Pi_D\,d^2\Sigma_D,          \label{eq:flops-env}    \\
    W_{\mathrm{sketch}} & \approx\ell\Pi_D\,(d^2\Sigma_D+\eta p), \label{eq:flops-sketch} \\
    W_{\mathrm{proj}}   & \approx r\Pi_D\,(d^2\Sigma_D+\eta p),   \label{eq:flops-proj}
  \end{align}
\end{subequations}
plus $\ord(\eta p\,\ell^2)$ for the QR\@. With $\eta=r=\ell$, a bulk site of the
right-to-left pass thus costs $2\ell\Pi_D(d^2\Sigma_D+\ell p)$. For an MPO--MPS product
($\Pi_D=D_1 D_\psi$, $p=d$), the term $\ell^2p\Pi_D$ reduces to the
$\ell^2d D_1 D_\psi$ of Ref.~\cite{camano2026src}, while $\ell\Pi_D d^2\Sigma_D$
accounts for absorbing the layers one by one and dominates when one input bond is much
larger than $\ell$. Keeping subleading powers of $d$ for an all-MPO stack, a contiguous
layer-by-layer path starting at the largest bond $D_{\max}$ costs
$\ell\Pi_D[d^2D_{\max}+d^3(\Sigma_D-D_{\max})]$ for the absorptions, plus $\ell\Pi_D p$
to close the physical legs in~\eqref{eq:ltr}. When several layers have comparable bond
dimensions, we use the exact flop count of the chosen contraction path.

Compressing a stack $A^{(1)}\cdots A^{(k)}$ in one sweep truncates once instead of
$k-1$ times, but $\Pi_D=\prod_{i}D_i$ in~\eqref{eq:flops} grows exponentially in the
depth, whereas pairwise application compresses each intermediate to $\chi$ and replaces
$\Pi_D$ by $\chi D_i$. In exchange, one sweep orthogonalizes each site once instead of
$k-1$ times, a saving that the leading terms understate because the QR runs far below
GEMM rates (\cref{sec:exp-quench}). A single sweep is therefore best suited to shallow
stacks of thin layers, such as a few Trotter layers or a Heisenberg step
$U^\dagger(\delta t)\,\hat O(t)\,U(\delta t)$~\cite{ouyang2026hubbard};
\cref{sec:exp-stacks,sec:exp-quench} test this trade-off empirically.

\paragraph{Synthetic performance benchmark.}
As a running sizing example and synthetic benchmark, we consider the stack
$A^{(1)}A^{(2)}A^{(3)}A^{(4)}$ of $k=4$ random MPOs on $n=50$ sites with physical
dimension $d=4$ ($p=16$) and bulk bond dimensions $D=(4,4,4000,4)$, compressed to
$\chi=\ell=2000$ in complex128. This mimics a large-bond operator dressed by shallow
gate layers, with $\Pi_D=4^3\cdot 4000=256\,000$ and $\Sigma_D=4012$. The cores of a bulk
site occupy $K\approx4000^2\cdot16\cdot16\,\mathrm{B}=4.1\GB$ (nearly all in $A^{(3)}$),
and the full 50-site stack occupies about $197\GB$. Unless stated otherwise, the cost
and memory estimates below refer to this benchmark in complex128; complex64 halves all
element-based byte counts.

Contraction-path selection strongly affects both memory and run time: a heuristic path
can create a large intermediate independent of the batch size or repeat work across
batches. Dynamic-programming path optimization~\cite{smith2018opteinsum} roughly halves
the sweep time at $D_3=1000$ on A100 GPUs (\cref{app:paths}) and is used by default,
with reported flop counts and rates taken from the executed paths.

\section{Parallel structure and the distributed sweep}\label{sec:parallel}

Every contraction in~\eqref{eq:ltr}--\eqref{eq:first} has an uncontracted batch index:
the sketch index $s$ in~\eqref{eq:ltr} and~\eqref{eq:sketch}, and the output bond
index $\rho$ in~\eqref{eq:project} and~\eqref{eq:first}. Partitioning along these
indices distributes the contractions and stored environments across GPUs
(\cref{sec:data-dist,sec:two-passes}) and bounds the working set on each device via
out-of-core batching (\cref{sec:ooc}), without changing the result beyond floating-point
rounding. Only the QR factorization couples the sketch columns, which a row-distributed
Cholesky QR reduces to small Gram-matrix exchanges (\cref{sec:dist-qr}). Unlike
site-partitioned fitting algorithms~\cite{fodera2026parallel}, this distribution
requires neither sub-chain updates nor a final merge, at the price of replicating the
projected environment (\cref{sec:comm}).

\begin{figure*}[t]
  \centering
  \begin{adjustbox}{max width=\linewidth}\input{figures/tikz/distributed_site.tex}\end{adjustbox}
  \caption{Distributed sweep with $G=4$.\ (a) Rank $g$ owns block $P_g$ of the sketch
    columns of every environment and block $R_g$ of the rows of every output core; the
    projected environment $S_{j+1}$ is replicated, and during the QR rank $g$ holds
    block $T_g$ of the rows of the sketch $M_j$.\ (b) One site of the right-to-left
    pass (schematic, not to scale): an all-to-all redistributes the sketch by rows, the
    Cholesky QR passes exchange only Gram matrices (thin bars), a second all-to-all
    distributes the rows of $\Gamma_j$, and each rank copies its rows to the host while
    it projects. The collectives make each site bulk-synchronous, and the first read of
    $C_j$ waits for the allgather of the previous site.}\label{fig:rtl}
\end{figure*}
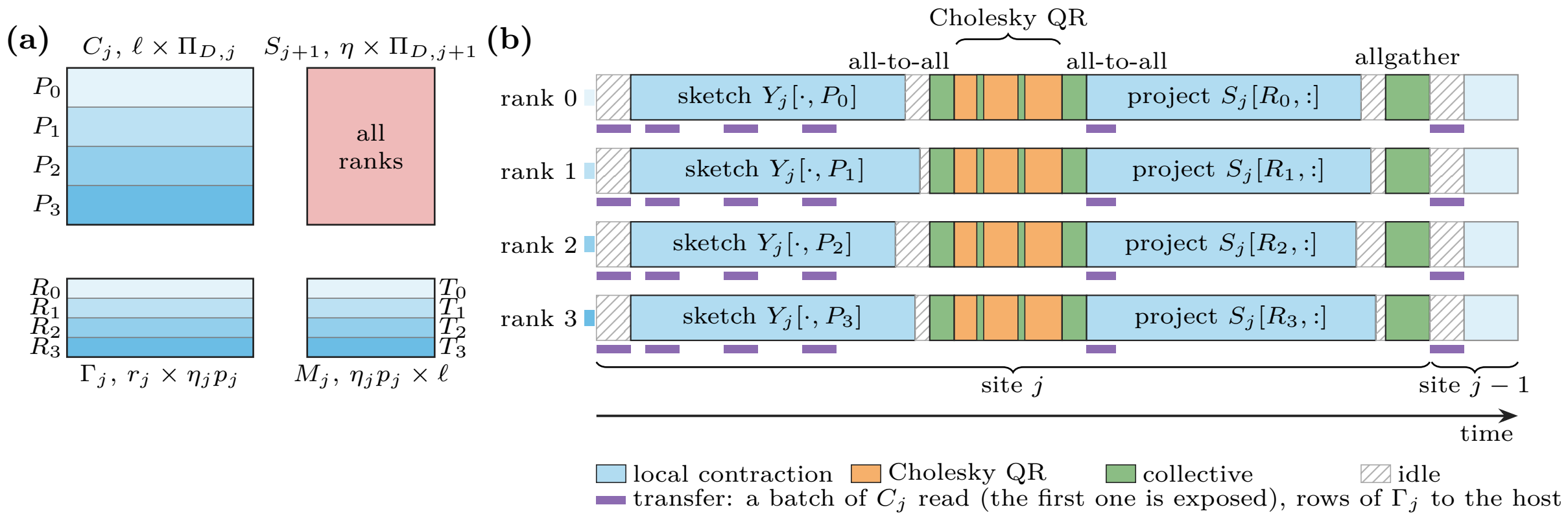

\subsection{Data distribution}\label{sec:data-dist}

A sweep across $G$ GPUs of a node uses $G$ ranks, $g\in\{0,\dots,G-1\}$, one per GPU,
executed by $G$ worker threads within a single host process rather than separate MPI
processes. Rank $g$
owns a contiguous block $P_g$ of the sketch columns, with block sizes differing by at
most one, and holds (\cref{fig:rtl}a):
\begin{itemize}[leftmargin=*]
  \item its slices $C_j[P_g,:]$ of every environment and $\Omega_j[P_g,:,:]$ of every
        sketch tensor;
  \item a full copy of the current projected environment $S_{j+1}$;
  \item during the QR, a block $T_g$ of the rows of the sketch $M_j$, with $\{T_g\}$ a
        balanced partition of its $\eta_{j}p_j$ rows;
  \item during the projection, a block $R_g$ of the rows of the output core
        $\Gamma_j$, with $\{R_g\}$ a balanced partition of $\{1,\dots,r_j\}$.
\end{itemize}
The environments, which dominate the memory, are thus divided among the ranks. Each
rank copies its rows of every output core to the host, where the cores
are assembled.

\subsection{The two passes}\label{sec:two-passes}

\paragraph{Left to right.} Each rank evaluates~\eqref{eq:ltr} on its own columns $s\in
P_g$; the environments need no communication.

\paragraph{Right to left.} At each site $j$ (\cref{fig:rtl,alg:sweep}):
\begin{enumerate}[leftmargin=*]
  \item each rank computes its columns $Y_j[\cdot,P_g]$ of the sketch from $C_j[P_g]$
        and the full $S_{j+1}$;
  \item an \textbf{all-to-all} redistributes the sketch from column blocks to row
        blocks: rank $g$ receives $M_j[T_g,:]$ with all $\ell$ columns;
  \item the ranks orthonormalize $M_j$ jointly by Cholesky QR (\cref{sec:dist-qr}),
        exchanging only Gram matrices, and rank $g$ obtains the rows $T_g$ of the
        isometry;
  \item a second \textbf{all-to-all} gives rank $g$ the rows $\Gamma_j[R_g]$ of the
        output core, which it copies to the host;
  \item each rank computes its rows $S_j[R_g,:]$ of~\eqref{eq:project};
  \item an \textbf{allgather} gives every rank the full $S_j$ for the next site.
\end{enumerate}
The first-site contraction~\eqref{eq:first} runs on rank 0. For $G=1$, the collectives
reduce to identities.

\begin{algorithm}[tb]
  \caption{Distributed SRC sweep, as executed by rank $g$.}\label{alg:sweep}
  \small
  \begin{algorithmic}[1]
    \STATE $C_1[P_g] \gets 1$
    \FOR{$j = 1$ \TO $n-1$}
    \STATE receive the cores of site $j$ \hfill\COMMENT{split upload, allgather}
    \FOR{each batch $B\subseteq P_g$}
    \STATE $C_{j+1}[B] \gets$~\eqref{eq:ltr}; store it \hfill\COMMENT{device, host or
    disk}
    \ENDFOR
    \ENDFOR
    \STATE $S_{n+1} \gets 1$
    \FOR{$j = n$ \TO $2$}
    \STATE receive the cores of site $j$; announce the reads of $C_j$
    \STATE $Y^{(g)} \gets$~\eqref{eq:sketch} over batches of $P_g$; release $C_j$
    \STATE $M_j[T_g,:] \gets \textsc{AllToAll}(Y^{(g)})$
    \STATE $Q_j[T_g,:] \gets$ Cholesky QR of $M_j$ \hfill\COMMENT{Gram allgathers}
    \IF{Cholesky QR declined}
    \STATE $\Gamma_j[R_g] \gets$ Householder QR \hfill\COMMENT{rank 0: gather, scatter}
    \ELSE
    \STATE $\Gamma_j[R_g] \gets \textsc{AllToAll}(Q_j[T_g,:])$
    \ENDIF
    \STATE copy $\Gamma_j[R_g]$ to the host
    \STATE $S_j[R_g] \gets$~\eqref{eq:project} over batches of $R_g$
    \STATE $S_j \gets \textsc{Allgather}(S_j[R_g])$
    \ENDFOR
    \IF{$g = 0$} \STATE $\Gamma_1 \gets$~\eqref{eq:first}; truncate if oversampled
    (\cref{sec:oversampling}) \ENDIF
  \end{algorithmic}
\end{algorithm}

\subsection{The distributed QR}\label{sec:dist-qr}

Cholesky QR~\cite{fukaya2014choleskyqr2} factorizes a tall matrix through its Gram
matrix: with $W=M_j^\dagger M_j=LL^\dagger$, the rows of $Q_j=M_{j}L^{-\dagger}$ can
be computed block by block. Each rank forms its partial Gram matrix
${M_j[T_g,:]}^\dagger M_j[T_g,:]$. The partial matrices are allgathered and summed in
rank order, so all ranks hold bitwise-identical sums and take identical decisions;
each rank then computes the Cholesky factor and solves for its rows. A single pass can
lose orthogonality proportionally to $u\,{\kappa(M_j)}^2$, where $u$ is the unit
roundoff and $\kappa(M_j)=\sigma_{\max}(M_j)/\sigma_{\min}(M_j)$ is the 2-norm
condition number. We therefore use three-pass shifted Cholesky
QR~\cite{fukaya2020shifted}: a first pass factorizes $W+\mu I$ with a shift $\mu$ of
order $(\eta_{j}p_j\ell+\ell^2)\,u\operatorname{tr}W$, followed by up to two unshifted
passes, requiring the Gram matrix entering the last pass to lie within $0.1$ of the
identity in the Frobenius norm (\cref{alg:cholqr} in \cref{app:cholqr}). With a cutoff,
every rank computes the SVD of the product $R_j$ of the triangular factors, of order
$\ell$, and truncates its rows of $Q_j$ by the rule of \cref{sec:rtl}. This reduces the
serial work per site from a Householder QR of $M_j$ ($16\eta_{j}p_j\ell^2$ real flops)
to two or three order-$\ell$ Cholesky factorizations replicated on every rank.

Early in an evolution without a cutoff, the bonds of an operator are padded to $\chi$
and its sketch is numerically rank-deficient. More generally, because each Gram pass
scales with $u\,{\kappa(M_j)}^2$, three-pass shifted Cholesky QR succeeds in double
precision ($u\approx1.1\times10^{-16}$) for moderately conditioned sketches, whereas in
single precision ($u\approx6.0\times10^{-8}$) physical sketches with rapidly decaying
singular values can exceed the conditioning threshold. When the Cholesky factorization
breaks down or the passes fail to reach the identity, all ranks detect the same
failure, and the site falls back to a Householder QR on rank 0, which gathers $M_j$,
factorizes it and scatters the rows of $\Gamma_j$. The fallback is also used for
$\ell<1024$, where the communication overhead exceeds the computational savings.

\subsection{Collectives within a node}

The collectives are peer-to-peer copies between device arrays, ordered
with respect to the kernels by CUDA events on one non-blocking stream per GPU\@. Events
mark both the publication of a source array and the completion of the receivers'
reads, so no buffer is reused while a copy is pending. Host barriers coordinate
publication, and alternating publication slots separate consecutive collectives. A
failure on one rank aborts the whole group instead of leaving the other ranks blocked.

\subsection{Reproducibility}

A single host generator draws the sketch tensors $\Omega_j$ in site order, and each
rank takes its own block of columns, making the sketch independent of the device count
and thread scheduling. In exact arithmetic, the compressed output is independent of
batching and partitioning whenever the QR step selects the same subspace. In
floating-point arithmetic, batch shapes and device counts change reduction orders,
contraction paths and QR sign conventions (or trigger the Householder fallback near the
conditioning threshold), and small differences can grow when singular values decay
slowly across the truncation point; we therefore compare full compressed tensors rather
than individual cores (\cref{sec:exp-scaling}).

\subsection{Communication volume}\label{sec:comm}

\Cref{tab:comm} lists the payload of each collective per site, with $e$ bytes per
element. The largest is the allgather of $S_j$, $(r_j-|R_g|)\Pi_{D,j}e$ bytes per
receiver, $6.1\GB$ for the synthetic benchmark with $G=4$; at an effective NVLink
bandwidth of $100\GB/\mathrm{s}$, it takes about $61\,\mathrm{ms}$, against tens of
seconds of computation per site. Over PCIe the same collective is ten times slower,
while the computation per site shrinks with $G$, so on eight GPUs it becomes the main
parallel overhead (\cref{sec:exp-scaling}). The QR exchanges are small: each
all-to-all and each Gram allgather moves $0.19\GB$ per receiver, a thirtieth of the
allgather of $S_j$.

\begin{table}[tb]
  \caption{Payload of the collectives at one site, excluding protocol overhead;
  $K_{j,g}\approx K_j/G$ is the share of the cores
    uploaded by rank $g$.}\label{tab:comm}
  \small
  \begin{tabular*}{\columnwidth}{@{\extracolsep{\fill}}llr@{}}
    \toprule
    Collective               & Bytes per receiver              & Benchmark \\
                             &                                 & ($G=4$)   \\
    \midrule
    all-to-all of $M_j$      & $|T_g|\,(\ell-|P_g|)\,e$        & $0.19\GB$ \\
    Gram allgather, per pass & $(G-1)\,\ell^2e$                & $0.19\GB$ \\
    all-to-all of $\Gamma_j$ & $|R_g|\,(\eta_{j}p_j-|T_g|)\,e$ & $0.19\GB$ \\
    allgather of $S_j$       & $(r_j-|R_g|)\,\Pi_{D,j} e$      & $6.1\GB$  \\
    core allgather, per pass & $K_j-K_{j,g}$                   & $3.1\GB$  \\
    \bottomrule
  \end{tabular*}
\end{table}

\section{Out-of-core execution}\label{sec:ooc}

The sweep must keep every sketched environment $C_j$ from the first pass until the
second pass consumes it. Their total size, $\ell e\sum_{j=2}^{n}\Pi_{D,j}$ bytes, grows
linearly with the chain length independently of $G$, and can exceed single-GPU or
single-node memory ($401\GB$ for the synthetic benchmark, versus $64\GB$ per A100 and
$512\GB$ of host RAM per Leonardo node). Out-of-core execution places each environment
on the device, in host memory or on disk within user-specified memory budgets; device-resident
objects such as $S_{j+1}$ and batch workspaces are listed in \cref{tab:memory} in
\cref{app:memory}.

\begin{figure*}[t]
  \centering
  \begin{adjustbox}{max width=\linewidth}\input{figures/tikz/ooc_pipeline.tex}\end{adjustbox}
  \caption{Data movement in the out-of-core sweep for one GPU of a node, with the
    thread or stream that drives each transfer and the nominal bandwidths of a
    Leonardo Booster node. Each environment $C_j$ resides in one tier: device memory,
    host memory or the scratch file system. The write and read rings hold two
    page-locked buffers each.}\label{fig:pipeline}
\end{figure*}
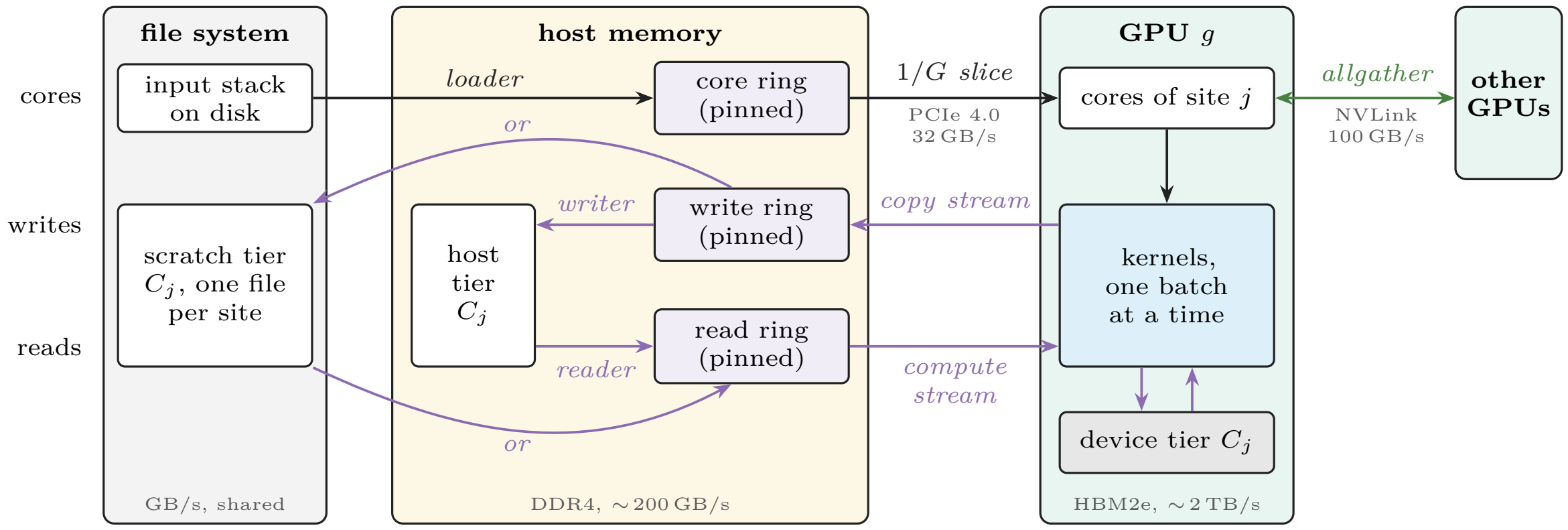

\subsection{The planner}\label{sec:planner}

The planner is a pure function of the core shapes, sketch size, data type, number of
devices and memory budgets, so a plan can be inspected before any data are read.
For each site and kernel, it selects the largest batch whose modelled peak memory
(cores, slices and the live intermediates along the chosen contraction path) fits the
device budget, rounded down to a multiple of 32 for GEMM efficiency. It then assigns
the environments to tiers newest-first: device, then host, then disk, because the
right-to-left pass reads the newest environments first. With $G$
devices, it plans for the largest column block and reserves on rank 0 the memory of
the Householder fallback, which gathers the full sketch, since any site may need it.

The memory pool can fragment beyond the margin reserved by the plan. A kernel
that runs out of memory then releases the cached blocks, retries with its batch halved
along the free index, and keeps the finer split for the rest of the sweep; since the
split index is free, the result is unchanged up to rounding.

\subsection{Storage tiers and pipelines}

Three pipelines move data between the tiers (\cref{fig:pipeline}):
\begin{enumerate}[leftmargin=*]
  \item \textbf{Cores.} A loader thread reads the cores in sweep order (sites
        $1,\dots,n-1$, then $n,\dots,1$) into a ring of page-locked buffers, one site
        ahead. With peer access, rank $g$ uploads about $1/G$ of each core over PCIe
        and allgathers the rest from its peers, so the host link carries every
        core once instead of $G$ times. Inputs can be memory-mapped files or chunked
        datasets.
  \item \textbf{Environment writes.} A completed batch is copied to a page-locked
        buffer on a separate copy stream, after an event recorded by the producing
        kernel; a writer thread then copies it into host memory or writes it to the
        scratch file.
  \item \textbf{Environment reads.} At the start of a site, its batches are announced
        and a reader thread fills two page-locked buffers ahead of use; the
        host-to-device copies are enqueued on the compute stream.
\end{enumerate}

\subsection{Overlap and exposed costs}\label{sec:overlap}

Moving environments off the device costs nothing as long as the pipelines keep up
with the kernels: in steady state, a batch must be stored or loaded in less time than
the GPU takes to process it, $t_{\mathrm{storage}}(b)\lesssim t_{\mathrm{compute}}(b)$.
The sweep satisfies this comfortably because its arithmetic intensity is
high~\cite{williams2009roofline}: each byte of an environment is used in many flops.
For the synthetic benchmark, a batch that takes about $2.4\s$ to process is read in
about $0.1\s$ from host memory and in under $1\s$ from the Lustre file system
(\cref{sec:exp-ooc}), so the environment traffic is hidden.

Host-to-device uploads on the compute stream add
$(n-1)\,|P_g|\,\Pi_D e/\beta_{\mathrm{PCIe}}$ per GPU during the right-to-left pass, about
$20\s$ on one GPU at $\beta_{\mathrm{PCIe}}=20\GB/\mathrm{s}$, under $1\%$ of the pass
time. Because reading $C_j$ begins only after the collectives at site $j+1$ finish,
each site also incurs a first-batch startup latency of $\sim0.1\s$ from host memory or
$\sim1\s$ from disk. Finally, loading input cores from a file system of bandwidth
$\beta_{\mathrm{fs}}$ takes $K_j/\beta_{\mathrm{fs}}$ per site in each pass regardless
of $G$, remaining hidden behind computation while
$K_j/\beta_{\mathrm{fs}}\lesssim t_{\mathrm{compute},j}(G)$. On Leonardo
($0.8$--$1.1\GB/\mathrm{s}$), reading $4.1\GB$ of cores takes $4$--$5\s$ per site
against $(16/G)\s$ of first-pass compute, reaching the file-system crossover at $G=4$
(\cref{sec:exp-scaling}).

\section{Performance model}\label{sec:model}

Let $H^{\mathrm{env}}_{j,g}$, $H^{\mathrm{sketch}}_{j,g}$ and $H^{\mathrm{proj}}_{j,g}$
be the real-flop counts of rank $g$ at site $j$, summed over the executed batches and
paths, and $F^{(\cdot)}_{j,g}$ the corresponding effective flop rates. Treating each
site as a bulk-synchronous superstep~\cite{valiant1990bsp}, we model the wall time on
$G$ GPUs as
\begin{equation}\label{eq:time}
  \begin{split}
    T_G \approx {} & \sum_{j=1}^{n-1}\max_g
    \frac{H_{j,g}^{\mathrm{env}}}{F_{j,g}^{\mathrm{env}}}
    +T_{\mathrm{first}}+T_{\mathrm{trunc}}+T_{\mathrm{data}}\\
    &+\sum_{j=2}^{n}\biggl[
      \max_g\frac{H_{j,g}^{\mathrm{sketch}}}{F_{j,g}^{\mathrm{sketch}}}
      +\max_g\frac{H_{j,g}^{\mathrm{proj}}}{F_{j,g}^{\mathrm{proj}}}\\
    &\hphantom{{}+\sum_{j=2}^{n}\biggl[}+T_{\mathrm{QR},j}+T_{\mathrm{coll},j}
      \biggr],
  \end{split}
\end{equation}
where $T_{\mathrm{QR},j}$ is the orthogonalization time at site $j$ (including Gram
exchanges) and $T_{\mathrm{coll},j}$ is the time spent in the remaining collectives at
site $j$, estimated from their byte volumes and link
bandwidths~\cite{thakur2005collectives}. The global terms $T_{\mathrm{first}}$,
$T_{\mathrm{trunc}}$ and $T_{\mathrm{data}}$ account for the first-site
contraction~\eqref{eq:first} on rank 0, the optional truncation sweep, and
non-overlapped host or disk transfers (\cref{sec:overlap}). When flops are evenly
balanced and serial work (site 1, truncation and order-$\ell$ Cholesky factorizations)
takes a fraction $f$ of the single-GPU time, Amdahl's law bounds parallel efficiency by
$E(G)\le 1/\bigl(1+(G-1)f\bigr)$ at fixed work and kernel rates. Plugging measured
single-GPU rates and link bandwidths into~\eqref{eq:time} gives a parameter-free strong-scaling
prediction~\cite{hoefler2015benchmarking} (\cref{sec:exp-scaling}).

Running Householder QR on rank 0 instead adds $16\eta p\ell^2$ real flops per site to
the serial work. For the synthetic benchmark this is $2\times10^{12}$ flops ($0.3\%$ of
the right-to-left work) and scarcely affects up to eight GPUs, whereas for thin stacks
it becomes significant: in \cref{sec:application} ($\eta p=4\ell$), rank-0 QR capped
four A100s at $1.6\times$ over one GPU at $\chi=2048$ in complex128, and distributed
Cholesky QR raised the speed-up to $2.3\times$. Because batching also alters path flop
counts (\cref{sec:flops}) and GEMM tile efficiency, measured efficiencies can slightly
exceed the idealized Amdahl bound.

\paragraph{Achieved rate.}
For the synthetic benchmark on one A100, the two passes sustain $16$ and $18.4\TFs$
(including non-overlapped I/O) from path flop counts and wall times, $75\%$ and $87\%$
of the $21.2\TFs$ of a large complex128 GEMM on this 124-SM part. In strict complex64 (TF32 disabled), device traces of
the right-to-left kernels measure $63.5\TFs$ on one RTX PRO 6000 ($76\%$ of its
$84\TFs$ large-GEMM rate) and $43.6\TFs$ on one GH200 ($79\%$ of its $55.4\TFs$
strict-FP32 GEMM rate; using its FP64 Tensor Cores, the GH200 reaches $64.1\TFs$ in
complex128). Run in complex128, the GH200 sweep of \cref{fig:scaling} takes $1357\s$ on
one GPU, only $18\%$ more than in complex64: double precision costs little compute there,
but it doubles the memory footprint.

\section{Numerical experiments}\label{sec:experiments}

\subsection{Setup}

Three hardware platforms serve complementary roles in the experiments: Leonardo Booster
nodes~\cite{leonardo_specs} test double-precision throughput and automatic Lustre
spilling, JUPITER Booster nodes~\cite{jupiter_specs} test strong scaling and
cost-model accuracy over high-bandwidth NVLink, and an eight-GPU PCIe node provides the
aggregate device ($760\,\mathrm{GiB}$) and host ($1.84\,\mathrm{TB}$) memory needed for
eight-GPU scaling and the 127-qubit evolutions up to $\chi=8192$.

Unless stated otherwise, double-precision measurements ran on Booster nodes of Leonardo
at CINECA\@. Each node has one 32-core Intel Xeon Platinum 8358 CPU, $512\GB$ of DDR4
RAM and four NVIDIA A100-SXM-64GB GPUs (a 124-SM variant, on which a large complex128
GEMM sustains $21.2\TFs$) fully connected by NVLink~3 (PCIe~4.0 to the host) without local disk, so inputs and spilled environments reside on Lustre scratch.
\code{portc} uses NumPy~\cite{harris2020numpy}, CuPy~\cite{okuta2017cupy} and
opt\_einsum~\cite{smith2018opteinsum}, with quimb~\cite{gray2018quimb} as the SVD and
approximate-contraction baseline (Python~3.14, NumPy~2.4.4, CuPy~14.2.0,
opt\_einsum~3.4.0, quimb~1.15.0, CUDA~13 on driver 535.274.02). GPU timings exclude
one-off kernel compilation.

The NVLink scaling benchmarks of \cref{sec:exp-scaling} ran on Booster nodes of JUPITER
at the J\"ulich Supercomputing Centre~\cite{jupiter_specs}. Each node comprises four
NVIDIA GH200 Grace Hopper Superchips, each pairing a 72-core Grace CPU ($120\GB$ of
LPDDR5X memory) with a Hopper GPU ($96\GB$ of HBM3, $95\,\mathrm{GiB}$ usable) over
NVLink-C2C ($900\GB/\mathrm{s}$), with full-mesh NVLink~4 between all four GPUs
($150\GB/\mathrm{s}$ per direction; concurrent peer collectives sustain
$130$--$133\GB/\mathrm{s}$ per receiving GPU; CUDA~13.2 on driver 610.57.04).

The out-of-core and PCIe scaling benchmarks of \cref{sec:exp-ooc,sec:exp-scaling} and
the circuit simulations of \cref{sec:application} ran on a cloud node with two 48-core
Intel Xeon 6776P CPUs, $1.84\,\mathrm{TB}$ of RAM and eight NVIDIA RTX PRO 6000
Blackwell Server Edition GPUs ($95\,\mathrm{GiB}$ each, four per socket on
PCIe~5.0~$\times16$, without NVLink or a shared PCIe switch; CUDA~13.2 on driver
580.173.02). Because FP64 throughput on the RTX PRO 6000 is low ($1.5\TFs$ in a
complex128 GEMM versus $84\TFs$ in complex64), the out-of-core and scaling runs on both
the PCIe and JUPITER nodes use complex64 with TF32 disabled on random unit-variance
stacks memory-mapped from a RAM disk. At $D_3=1000$, switching from complex128 to
complex64 for the same stack and seed changes the output by $6.4\times10^{-5}$ in
relative Frobenius norm while speeding up the sweep 27-fold on the RTX PRO 6000.

Unless stated otherwise, $\ell=\chi$, the relative cutoff is zero, and flop rates count
real operations, with eight flops per complex multiply--add. Operator errors are
relative Frobenius errors, the state infidelity is
$1-|\langle\psi,\phi\rangle|^2/(\|\psi\|_2^2\|\phi\|_2^2)$, and observable errors are
absolute differences.

\begin{table}[tb]
  \caption{MPO--MPO product and compression, complex128, one Leonardo node (32 cores,
    one A100); $\chi_{\mathrm{id}}=4$ at $\chi=1000$. Speed-up of SRC on the GPU over
    quimb on the CPU.}\label{tab:primitives}
  \small
  \setlength{\tabcolsep}{4pt}
  \begin{tabular*}{\columnwidth}{@{\extracolsep{\fill}}rrrrrr@{}}
    \toprule
        &        & quimb   & \multicolumn{2}{c}{SRC} &          \\
    \cmidrule(lr){3-3}\cmidrule(lr){4-5}
    $n$ & $\chi$ & CPU (s) & CPU (s) & GPU (s)       & Speed-up \\
    \midrule
    50  & 50     & 37.76   & 0.23    & 0.57          & 66       \\
    20  & 100    & 17.86   & 0.25    & 0.46          & 39       \\
    25  & 1000   & 422.57  & 60.47   & 2.32          & 182      \\
    50  & 1000   & 1040.72 & 137.37  & 4.47          & 233      \\
    \bottomrule
  \end{tabular*}
\end{table}

\subsection{MPO--MPO products}\label{sec:exp-primitives}

We multiply two MPOs and compress the product back to the input bond dimension $\chi$.
One MPO is random and the other a perturbed identity of bond dimension
$\chi_{\mathrm{id}}$, so that the product is highly compressible. \Cref{tab:primitives}
compares SRC with quimb's contract-then-compress (randomized
SVD) on the same
node. At small $\chi$ the GPU is underutilized and the CPU is faster; at $\chi=1000$
the GPU is $182$--$233$ times faster than quimb on 32 cores, and SRC needs one seventh
of quimb's host memory. In complex64, the GPU times at $\chi=1000$ drop to $1.90$ and
$3.03\s$.

\subsection{Single-sweep versus pairwise compression}\label{sec:exp-stacks}

We compare compressing a stack in a single sweep,
$\operatorname{SRC}(A^{(1)},\dots,A^{(k-1)},\psi)$, against pairwise application that
truncates to $\chi$ after every product. On 10-site chains, we measure the median
relative error against the exact dense product over 20 random instances, alongside the
largest best rank-$\chi$ truncation error across all bonds, a lower bound for any train
of bond dimension $\chi$. On 30-site chains with $\chi=64$, we measure the run time.

\paragraph{Accuracy.}
One sweep is at best moderately more accurate. For random stacks terminated by an MPS,
it lowers the median error by about $20\%$ at depth 4 and beats pairwise application in
$80$--$90\%$ of the instances; for random MPO products the gain is at most $8\%$, and
for Trotter layers there is no systematic difference. Both methods lie 2--15 times
above the lower bound, so eliminating the intermediate truncations alone does not close
the gap to the best fixed-rank approximation.

\paragraph{Cost.}
As predicted by~\eqref{eq:flops}, thin Trotter layers (bond dimension at most 4) take
$0.6$--$1.4$ times the pairwise time up to depth 5, whereas random MPOs of bond
dimension 8 and 16 at depth 4 are $30$ and $136$ times slower (\cref{tab:depth} in
\cref{app:benchmarks}).

\subsection{Fusing Trotter steps}\label{sec:exp-quench}

To test stack fusion on thin physical layers, we evolve mixed-field Ising chains of 50
and 100 qubits, $H=\sum_{i}Z_{i}Z_{i+1}+1.05\sum_{i}X_i+0.5\sum_{i}Z_i$, at the
non-integrable point of Ref.~\cite{banuls2011thermalization}, from $|0\cdots0\rangle$
to $t=8$ in 80 first-order Trotter steps of $\delta t=0.1$ (each an MPO of bond
dimension two). On one A100, with dynamic-programming paths, SRC applies $m$ steps per
sweep as a stack of depth $k=m+1$. Errors in \cref{tab:quench} are measured relative to
the $m=1$, $\chi=1024$ evolution, for which the 50- and 100-qubit chains agree on the
central $\langle Z\rangle$ to $10^{-6}$.

Fusing two or four steps is $1.5$--$2.7$ times faster than applying one step at a time
and increases the infidelity by at most $8\%$. Compared with the 32-core quimb SVD
baseline, SRC has higher infidelity at equal $\chi$ ($0.062$ versus $0.037$ at
$\chi=64$), whereas doubling the bond dimension to $\chi=128$ with $m=4$ lowers the
infidelity to $0.022$ in $5.7\s$.

Fusion speeds up the evolution by doing fewer QR factorizations and fewer state
transfers, not by reducing the contraction work. Substituting $\Pi_D=2^m\chi$
into~\eqref{eq:flops}, the leading contraction cost of one fused sweep is
$2^{m-1}/m$ times that of applying the same $m$ steps one at a time. The cost is
unchanged for two steps and doubles for four. At $\chi\ge512$, the contractions
sustain $9$--$18\TFs$ with or without fusion, or $40$--$83\%$ of the $21.2\TFs$
measured for a large complex128 GEMM on this GPU. In both cases, their arithmetic
intensity is above the roofline ridge point~\cite{williams2009roofline}
of $15\,\mathrm{flop/B}$.

With fusion, we perform the Householder QR of each sketch and transfer the state
between host and device only once every $m$ steps. The QR runs at $0.3$--$2.4\TFs$
and accounts for $36$--$45\%$ of the pairwise run time. In the instrumented run at
$\chi=512$, fusing two steps saves $16\s$ relative to applying one step at a time.
About half of that saving comes from QR, and nearly all of the rest comes from
transfers and host overhead (\cref{fig:quench-breakdown} in \cref{app:benchmarks}).
At the same $\chi$, fusing four steps is slower than fusing two: the extra
contraction work costs more time than the additional QR savings.

\begin{table}[tb]
  \caption{Quench of the mixed-field Ising chain to $t=8$ on one A100, fusing $m$
  Trotter steps per sweep; quimb runs on 32 cores.}\label{tab:quench}
  \small
  \setlength{\tabcolsep}{4pt}
  \begin{tabular*}{\columnwidth}{@{\extracolsep{\fill}}rrlrrr@{}}
    \toprule
    $n$ & $\chi$ & Method     & Time (s) & Infidelity         & $\Delta\langle Z_{n/2}\rangle$ \\
    \midrule
    50  & 64     & quimb SVD  & 739      & 0.0368             & 0.0041                         \\
    50  & 64     & SRC, $m=1$ & 10.9     & 0.0616             & 0.0134                         \\
    50  & 64     & SRC, $m=4$ & 4.88     & 0.0631             & 0.0135                         \\
    50  & 128    & SRC, $m=4$ & 5.68     & 0.0218             & 0.0060                         \\
    50  & 512    & SRC, $m=1$ & 37.0     & $6.6\times10^{-4}$ & $2.4\times10^{-4}$             \\
    50  & 512    & SRC, $m=2$ & 21.8     & $6.9\times10^{-4}$ & $2.6\times10^{-4}$             \\
    50  & 512    & SRC, $m=4$ & 24.9     & $7.1\times10^{-4}$ & $2.8\times10^{-4}$             \\
    100 & 256    & SRC, $m=1$ & 33.3     & 0.0131             & 0.0017                         \\
    100 & 256    & SRC, $m=4$ & 15.5     & 0.0137             & 0.0018                         \\
    100 & 512    & SRC, $m=1$ & 76.1     & $1.8\times10^{-3}$ & $2.5\times10^{-4}$             \\
    100 & 512    & SRC, $m=2$ & 45.0     & $1.8\times10^{-3}$ & $2.5\times10^{-4}$             \\
    \bottomrule
  \end{tabular*}
\end{table}

\begin{table*}[t]
  \caption{Storage tiers for the synthetic benchmark on one RTX PRO 6000 in complex64:
    tiers of the 50 environment slots (device/host/disk, including the trivial
    boundary slot), planned device peak and final memory-pool size, planned host peak
    and maximum resident set size (RSS, including mapped input pages), data spilled,
    and wall times of the sweep and its two passes. The disk is virtual, at
    $0.15\GB/\mathrm{s}$.}\label{tab:ooc}
  \small
  \begin{tabular*}{\textwidth}{@{\extracolsep{\fill}}lrrrrrr@{}}
    \toprule
    Plan                & Tiers  & Device (GB) & Host (GB)     & Spilled & Wall   & L$\to$R / R$\to$L \\
                        & d/h/s  & plan / pool & plan / RSS    & (GB)    & (s)    & (s)               \\
    \midrule
    default             & 7/43/0 & 91.1 / 81.7 & 223.6 / 298.1 & 0       & 842.3  & 238.6 / 601.2     \\
    GPU budget $16\GB$  & 1/49/0 & 15.9 / 22.8 & 246.1 / 318.8 & 0       & 799.3  & 207.4 / 589.2     \\
    host budget $50\GB$ & 7/0/43 & 91.1 / 81.7 & 47.5 / 144.0  & 176.1   & 2859.1 & 2255.1 / 600.8    \\
    \bottomrule
  \end{tabular*}
\end{table*}

\subsection{Out-of-core execution}\label{sec:exp-ooc}

On the PCIe node, we vary the planner budgets to exercise all three storage tiers for
the synthetic benchmark in complex64 (\cref{tab:ooc}). Lowering the device budget to
$16\GB$ moves nearly all environments to host memory without a run-time penalty.
Spilling to the $0.15\GB/\mathrm{s}$ virtual disk increases the wall time $3.4$-fold
through slow writes in the left-to-right pass, while reads in the right-to-left pass
remain hidden behind computation. Plans that differ only in storage placement return
bit-identical outputs, and allocator caching can leave the final memory pool slightly
above the planned live-memory budget (\cref{sec:planner}).

On Leonardo in complex128, disk spilling occurs automatically: although the $401\GB$ of
environments fit within the nominal host budget ($90\%$ of $512\GB$), adding output and
staging buffers exceeds it, spilling 3 to 14 environments to Lustre
($0.8$--$1.1\GB/\mathrm{s}$). During the first right-to-left sketch, pool fragmentation
triggered the automatic batch halving of \cref{sec:planner} in every multi-GPU run and
one of two single-GPU runs, after which all sweeps completed within $3\%$ of their
planned host peaks.

\subsection{Multi-GPU scaling}\label{sec:exp-scaling}

\begin{figure*}[tbp]
  \centering
  \includegraphics[width=\textwidth]{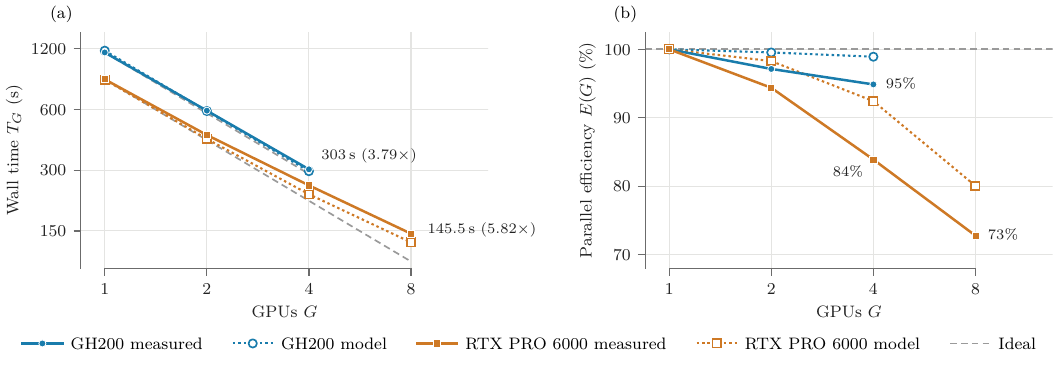}
  \caption{Strong scaling of the 50-site synthetic benchmark ($D_3=4000$, $\chi=2000$,
    strict complex64, dynamic-programming contraction paths) on a four-GPU JUPITER
    GH200 node (NVLink~4) and on the eight-GPU RTX PRO 6000 node (PCIe~5.0).\ (a) Wall
    time $T_G$; dashed lines show ideal $T_1/G$ scaling from the measured one-GPU
    baselines ($1150.7\s$ on GH200, $847.0\s$ on RTX PRO 6000), and dotted curves show
    pre-run predictions of~\eqref{eq:time} from a single traced one-GPU sweep and
    standalone link microbenchmarks.\ (b) Parallel efficiency $E(G)=T_1/(GT_G)$, with
    each curve normalized by its own $G=1$ run time.}\label{fig:scaling}
\end{figure*}

\Cref{fig:scaling} compares strong scaling of the 50-site synthetic benchmark
($D_3=4000$, $\chi=2000$, strict complex64) on a four-GPU JUPITER Booster node
(NVLink~4) and on the eight-GPU PCIe~5.0 node. Because both GPUs expose
$95\,\mathrm{GiB}$ of device memory, the planner selects identical environment tiers on
both platforms ($7$, $15$ and $33$ of the $50$ slots on device at $G=1,2,4$, and all
$50$ at $G=8$). On JUPITER, the median wall time across two runs drops from $1150.7\s$
on one GH200 to $592.4\s$ on two and $303.2\s$ on four, corresponding to speed-ups of
$1.94\times$ and $3.79\times$ and parallel efficiencies of $97\%$ and $95\%$ (one GH200
is slower than one RTX PRO 6000 here because strict complex64 bypasses Tensor Cores;
\cref{sec:model}). On the PCIe node, the wall time drops from $847.0\s$ on one GPU
(median of two runs) to $448.7$, $252.4$ and $145.5\s$ on two, four and eight GPUs,
giving speed-ups of $1.89\times$, $3.36\times$ and $5.82\times$ and efficiencies of
$94\%$, $84\%$ and $73\%$.

Before running any multi-GPU sweeps on either platform, we evaluated the performance
model~\eqref{eq:time} using only a single traced one-GPU sweep and seconds-long
standalone microbenchmarks of peer-copy bandwidths and the order-$\ell$ QR kernel, with
no fitted parameters. On JUPITER, this pre-run calculation predicted wall times of
$590.0\s$ on two GPUs and $296.8\s$ on four, matching the measured untraced medians
within $0.4\%$ and $2.1\%$ ($1.1\%$ and $2.3\%$ against traced multi-GPU runs; in
\cref{fig:scaling}b, normalizing each curve by its own $G=1$ baseline includes the $2\%$
tracing overhead in $T_1^{\mathrm{traced}}=1174.5\s$, giving predicted efficiencies of
$99.5\%$ and $98.9\%$). Rank-0 device traces confirm why~\eqref{eq:time} is so accurate
over NVLink: right-to-left contraction time halves cleanly with $G$ ($844.5$, $414.1$
and $207.9\s$ at $G=1,2,4$, sustaining $\approx44\TFs$ per GPU), while all collectives
and distributed Cholesky QR together take $10.8\s$ ($4.9\%$ of the pass) at $G=4$. When
the interconnect delivers its probed bandwidth, the bulk-synchronous decomposition in
\eqref{eq:time} accounts for the multi-GPU run time directly from single-GPU
measurements, showing little unmodelled synchronization or contention.

On the PCIe node, the same pre-run procedure predicted efficiencies of $98\%$, $92\%$
and $80\%$, exceeding the measurements by a margin that grows with $G$. Because
\eqref{eq:time} separates local contractions from collective transfers ($H/F$ versus
$T_{\mathrm{coll},j}$), comparing each term against GPU traces isolated the cause. In a
traced first version that issued all kernels and copies on the default CUDA stream
($63\%$ efficiency at $G=8$), kernel execution scaled as modelled, keeping total GPU
compute time constant across $G$, whereas the allgathers of $S_{j+1}$ and input cores
achieved under half the probed bandwidth on eight GPUs: peer copies serialized across
destinations on the default stream, and without a PCIe switch every transfer traversed
the dual-socket CPU root complexes. Assigning each GPU a dedicated non-blocking stream
raised eight-GPU efficiency to the $73\%$ shown in \cref{fig:scaling}.

The PCIe scaling runs predated the distributed Cholesky QR of \cref{sec:dist-qr} and
therefore used the rank-0 Householder QR path, whereas the JUPITER runs used
distributed Cholesky QR (accepted in complex64 at nearly every site on these
well-conditioned random stacks, leaving $0.2\s$ in Householder fallbacks). At
$D_3=4000$, this code difference accounts for very little of the efficiency gap between
the two machines: in the PCIe traces, rank-0 Householder QR ($2.7\s$) plus its sketch
gather and core scatter ($1.3\s$ at $G=4$, $1.6\s$ at $G=8$) total $4.0\s$ ($1.6\%$ of
wall time) on four GPUs and $4.3\s$ ($3.0\%$) on eight, and distributed Cholesky QR
still exchanges the same byte volume in two all-to-alls (\cref{tab:comm}). Instead, the
higher four-GPU efficiency on JUPITER ($95\%$ versus $84\%$) reflects its tenfold higher
peer bandwidth on the $S_{j+1}$ and input-core allgathers.

Shrinking the large bond to $D_3=1000$ ($\chi=\ell=500$) reduces per-site compute to
$35$--$45\,\mathrm{ms}$ per GPU and puts both platforms on the same rank-0 Householder
QR path, since $\ell<1024$ (\cref{sec:dist-qr}). On JUPITER, median wall time still
scales from $23.4\s$ on one GH200 to $11.5\s$ on two ($101.5\%$ efficiency, aided by
finer batch paths) and $6.43\s$ on four ($3.63\times$, $91\%$ efficiency); over PCIe,
efficiency drops to $66\%$ at $G=4$ ($20.3\s$ to $7.6\s$) and eight GPUs run slower than
four ($9.7\s$). On Leonardo's NVLink-connected A100s in complex128, the full $D_3=4000$
benchmark runs in $3925\s$ on one GPU and $1100\s$ on four ($89\%$ efficiency), while a
second four-GPU run took $1290\s$ owing to longer input-read stalls on Lustre at the
file-system crossover of \cref{sec:overlap}. Across partitions, multi-GPU outputs on
both JUPITER and the PCIe node differ from the one-GPU output by
$1.9$--$2.7\times10^{-4}$ in relative Frobenius norm at $D_3=4000$
($6.6$--$8.5\times10^{-5}$ at $D_3=1000$), matching the shift seen on one GPU when
varying only the batch size on a spectrum that decays slowly across $\chi$.

\section{Application: a 127-qubit kicked Ising circuit}\label{sec:application}

We apply the full pipeline to a standard workload with published reference values: the
Heisenberg-picture evolution of local observables under the 127-qubit kicked Ising
circuit of Kim et al.~\cite{kim2023utility}. Classical simulations of this circuit
include belief propagation on the native heavy-hex tensor
network~\cite{tindall2024eagle}, sparse Pauli dynamics~\cite{begusic2024converged} and
Heisenberg-picture MPOs~\cite{anand2023zne}. We use it here to benchmark stacked SRC
against published 1D and 2D references and to exercise the distributed, out-of-core
sweep at large operator bond dimensions.

\subsection{Circuit and stacks}\label{sec:app-setup}

\begin{figure}[tbp]
  \centering
  \begin{adjustbox}{max width=\linewidth}\input{figures/tikz/kicked_stack.tex}\end{adjustbox}
  \caption{The kicked Ising network in the Heisenberg picture (green: gate sub-layers
    $U_s$, six per Trotter step; blue: the observable $O$, initially $Z_{62}$).\ (a)
    One SRC sweep compresses the stack $U_s^\dagger O U_s$ (depth $k=3$, uncompressed
    bond dimension $4\chi$) to bond dimension $\chi$; later sweeps absorb the remaining
    sub-layers (faded).\ (b) Schematic 8-qubit fragment of one sub-layer $U_s$: because
    its $R_{ZZ}$ gates do not overlap along the 1D chain, $U_s$ is an MPO of bond
    dimension at most 2.}\label{fig:kicked-stack}
\end{figure}
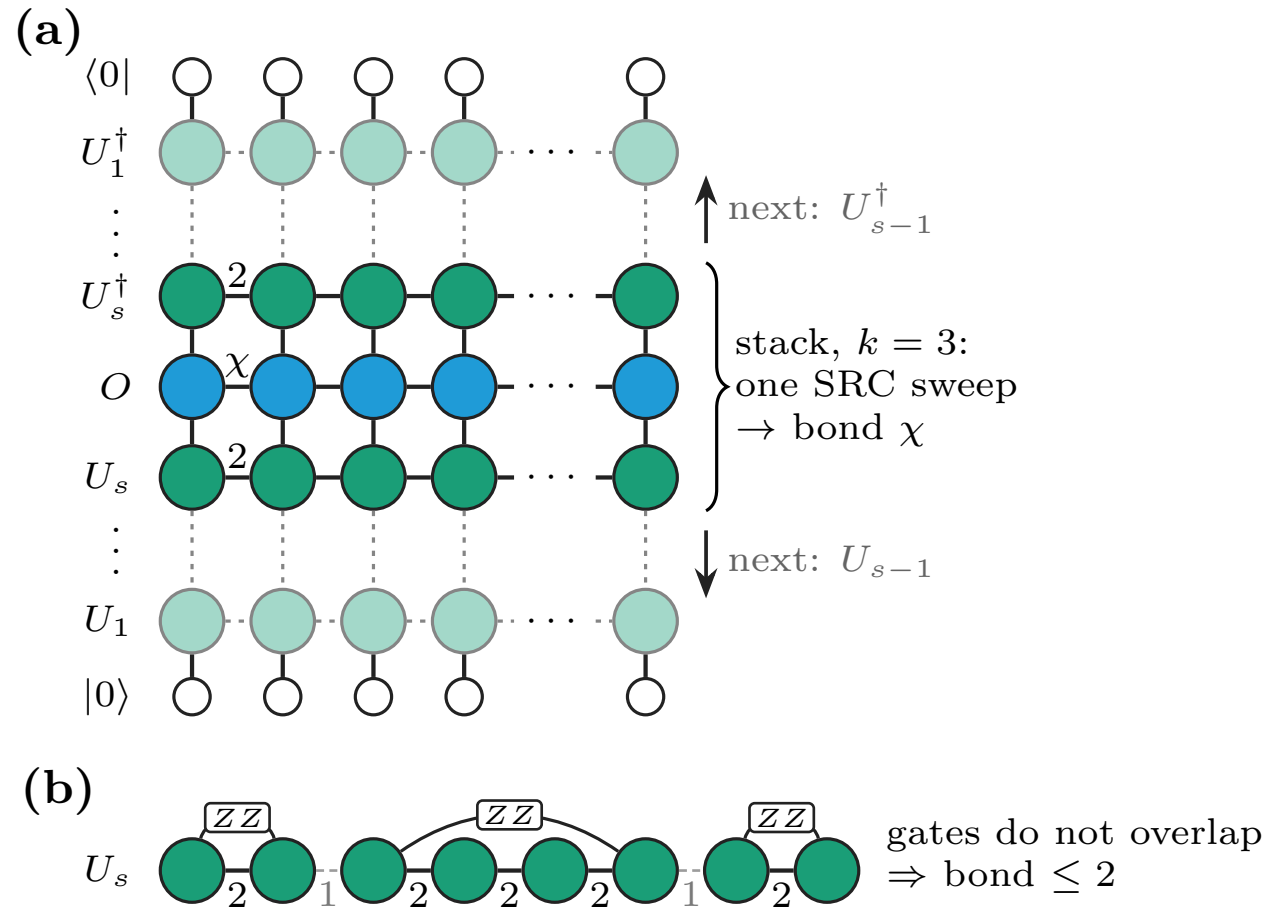

Each Trotter step applies $R_X(\theta_h)$ to every qubit, followed by $R_{ZZ}(-\pi/2)$
on the 144 edges of the 2D heavy-hex coupling map of the device, with
$R_X(\theta)=\exp(-i\theta X/2)$ and $R_{ZZ}(\theta)=\exp(-i\theta Z\otimes Z/2)$. We
order the qubits along a 1D chain by their hardware labels, as in
Refs.~\cite{kim2023utility,anand2023zne}. In this ordering, a two-qubit gate between
qubits $i<j$ spans every bond of the chain between them, and up to 6 of the 144 gates
of a Trotter step cross the same chain bond. Since the $R_{ZZ}$ gates commute, we
partition the 144 gates of each Trotter step into 6 sub-layers $U_s$ whose gates never
overlap along the chain, the minimum for this ordering (Anand et al.\ used 13). Every
sub-layer $U_s$ (and $U_s^\dagger$) is then a single MPO of bond dimension at most 2
(\cref{fig:kicked-stack}b), with the single-qubit $R_X$ rotations absorbed into the
last sub-layer of the step; a 20-step evolution thus consists of $20\times6=120$
sub-layers.

In the Heisenberg picture, we evolve an observable $O$ as an MPO from the end of the
circuit to the beginning by applying one sub-layer at a time,
$O\leftarrow U_s^\dagger O U_s$, and finally contracting the evolved MPO with
$|0\rangle^{\otimes 127}$. Each sub-layer update is one stack
$\operatorname{SRC}_\chi(U_s^\dagger, O, U_s)$ of depth $k=3$, with uncompressed
bond dimension $2\times\chi\times2=4\chi$ and physical dimension $2\times2$ per site,
compressed in one sweep without an intermediate truncation (\cref{fig:kicked-stack}a);
fusing $m$ sub-layers into a single stack of depth $k=2m+1$ raises the uncompressed
bond dimension to $4^m\chi$.

All runs in this section use the PCIe node of \cref{sec:experiments}, with one GPU per
run unless stated otherwise. The validation and comparisons of
\cref{sec:app-validation,sec:app-compression} use complex128. The long evolutions of
\cref{sec:app-scale,sec:app-physics} use complex64, which at $\chi=1024$ runs 16 times
faster ($23.5$ versus $372\s$ per step) and shifts the 20-step $\langle Z_{62}\rangle$
by $1.9\times10^{-3}$. That shift is well below the $\approx 0.08$ discrepancy with the
published mixed-picture 2D tensor-network (MIX) benchmark~\cite{begusic2024converged},
which evolves the state forward and the observable backward for half the circuit each
on the heavy-hex graph (\cref{sec:app-physics}). Because the tested quimb routines were
unstable or much slower in complex64, all quimb runs use complex128.

\subsection{Validation}\label{sec:app-validation}

On a 12-qubit heavy hexagon, all tested methods reproduce the dense $U^\dagger O U$ to
$10^{-14}$ in relative Frobenius norm, and on a 16-qubit fragment the expectation
values agree with a state-vector simulation to $4\times10^{-15}$. At the Clifford
points $\theta_h=0$ and $\theta_h=\pi/2$, the evolved observables remain single Pauli
strings; on all 127 qubits, $\langle Z_{62}\rangle$ after 20 steps and the weight-10
and weight-17 observables of Ref.~\cite{kim2023utility} after 5 steps match a
stabilizer simulation~\cite{gidney2021stim} to $10^{-12}$. Away from the Clifford
points, the 5-step weight-10 observable matches the exact values of
Ref.~\cite{begusic2024converged} to $10^{-8}$ across their angle grid for $\chi\ge 384$
(where Anand et al.\ also find it exact~\cite{anand2023zne}), and to $2\times10^{-12}$
for $\chi\ge 512$. At $\chi=1024$ and $\theta_h=8\pi/32$, its error is
$2.9\times10^{-6}$ in complex64 versus $8\times10^{-16}$ in complex128. The 5-step
weight-17 observable at $\chi=1024$ agrees with the exact values within
$1.2\times10^{-4}$, except at $\theta_h=10\pi/32$ ($4.0\times10^{-4}$) and $11\pi/32$
($9.1\times10^{-4}$).

\subsection{Quality of a single compression}\label{sec:app-compression}

\Cref{tab:app-compression} isolates a single compression: we evolve $Z_{62}$ for 8
steps at $\theta_h=6\pi/32$, apply the first sub-layer of step 9, and compress the
$4\chi$-bond stack back to $\chi$. At $\chi=256$, \code{portc} is $2.3$ times more
accurate than quimb's SRC in nearly the same time and shows smaller seed-to-seed
spread (three 9-step runs agree to $10^{-3}$ in $\langle Z_{62}\rangle$, versus
$2\times10^{-2}$ for quimb). In quimb~1.15.0, the SRC routine draws real
(\code{float64}) Gaussian sketches even for complex networks and truncates during the
right-to-left pass, whereas \code{portc} draws complex Gaussian sketches and defers
truncation to the final canonical SVD pass when oversampling (\cref{sec:oversampling}).
With oversampling, \code{portc} achieves the lowest error among the approximate
contraction methods at $\chi=256$, ahead of zip-up and variational fitting; at
$\chi=64$, variational fitting is slightly more accurate at eight times the run time.
The deterministic TT-SVD baseline~\cite{oseledets2011tt} reaches lower errors at both
bond dimensions ($0.048$ and $0.033$) at $1.5$--$2.8$ times the cost of oversampled
SRC\@.

\begin{table}[tb]
  \caption{Single compression of $U_s^\dagger O U_s$ (uncompressed bond $4\chi$) back
  to $\chi$, for $Z_{62}$ after 8 steps at $\theta_h=6\pi/32$ and the
    first sub-layer of step 9. Error: relative Frobenius error of the compressed
    operator, mean over 5 seeds for the randomized methods; one RTX PRO 6000,
    complex128. Oversampled rows use $\ell=\max(\lceil1.5\chi\rceil,\chi+10)$ and a
    final SVD truncation. The variational fit uses a cutoff of
    $10^{-12}$.}\label{tab:app-compression}
  \small
  \setlength{\tabcolsep}{4pt}
  \begin{tabular*}{\columnwidth}{@{\extracolsep{\fill}}lrrrr@{}}
    \toprule
                          & \multicolumn{2}{c}{$\chi=64$} & \multicolumn{2}{c}{$\chi=256$}                    \\
    \cmidrule(lr){2-3}\cmidrule(l){4-5}
    Method                & Error                         & Time (s)                       & Error & Time (s) \\
    \midrule
    TT-SVD                & 0.048                         & 2.0                            & 0.033 & 16.8     \\
    SRC, this work        & 0.115                         & 0.5                            & 0.080 & 2.3      \\
    \quad oversampled     & 0.073                         & 1.3                            & 0.051 & 6.0      \\
    quimb SRC             & 0.269                         & 0.7                            & 0.187 & 2.3      \\
    \quad oversampled     & 0.151                         & 1.5                            & 0.106 & 5.9      \\
    quimb zip-up          & 0.080                         & 2.1                            & 0.056 & 12.9     \\
    quimb variational fit & 0.066                         & 10.5                           & 0.116 & 34.9     \\
    \bottomrule
  \end{tabular*}
\end{table}

Over a full evolution at moderate $\chi$, zip-up with a left-canonical input reaches a
target error at smaller $\chi$ and in less time for the 9-step $\langle Z_{62}\rangle$
and 5-step weight-17 observable at $\theta_h=8\pi/32$ and $10\pi/32$, while at
$6\pi/32$ the two methods perform similarly ($2.2\times10^{-3}$ in $330\s$ for zip-up
versus $2.5\times10^{-3}$ in $657\s$ for SRC). Fusing multiple sub-layers does not pay
off here: at fixed $\chi$, errors for $m=2$ and $3$ ($k=5$ and $7$) stay within a
factor of three of $m=1$ ($k=3$) without a consistent trend, while $m=3$ runs 3--5
times slower and uses 7--15 times more device memory. Because truncating after a single
gate per bond loses little accuracy, fusion has little error to avoid
(\cref{sec:flops,sec:exp-stacks}).

\begin{table*}[t]
  \caption{The 20-step evolution of $\langle Z_{62}\rangle$ at $\theta_h=8\pi/32$ with
    SRC ($m=1$ sub-layer per sweep, $k=3$, cutoff 0) on the PCIe node: median time per
    step, total time, aggregate flop rate across all active GPUs, final size of the
    memory pool per GPU, and host peak. The operator occupies $3.9$, $15.6$, $62$ and
    $245\GB$ in complex64 at the four bond dimensions. The eight-GPU runs use the
    Householder QR on rank 0.}\label{tab:app-cost}
  \small
  \begin{tabular*}{\textwidth}{@{\extracolsep{\fill}}rrlrrrrr@{}}
    \toprule
    $\chi$ & GPUs & Precision  & s/step & 20 steps (h) & Total $\mathrm{TFLOP/s}$ & Device (GB) & Host (GB) \\
    \midrule
    1024   & 1    & complex128 & 372    & 2.05         & 1.1                      & 9.4         & 32        \\
    1024   & 1    & complex64  & 23.5   & 0.13         & 17.7                     & 4.7         & 13        \\
    2048   & 1    & complex64  & 106    & 0.58         & 31.5                     & 18.8        & 49        \\
    4096   & 8    & complex64  & 404    & 2.19         & 67.0                     & 13.0        & 212       \\
    8192   & 8    & complex64  & 1776   & 9.74         & 120.0                    & 49.2        & 916       \\
    \bottomrule
  \end{tabular*}
\end{table*}

\subsection{Large bond dimensions}\label{sec:app-scale}

After a few steps, the operator saturates $\chi$ across most of the chain, so each
subsequent sub-layer compresses a full 127-site stack of uncompressed bond dimension
$4\chi$. \Cref{tab:app-cost} reports the cost of the 20-step evolution of $Z_{62}$ at
$\theta_h=8\pi/32$ from $\chi=1024$ to $8192$. To our knowledge, $\chi=8192$ is the
largest bond dimension reported for a 1D tensor-network simulation of the full 127-qubit
circuit; eight times the $\chi=1024$ reached with Heisenberg MPOs by Anand et
al.~\cite{anand2023zne} and with light-cone MPS by Kim et al.~\cite{kim2023utility}
(who used larger $\chi$ only on reduced circuits). At $\chi=8192$, a single operator
occupies $245\GB$ in complex64. On eight GPUs, the 20-step run stays in host and device
memory without disk spilling ($49\GB$ per GPU and $916\GB$ on the host) and finishes in
$9.7\,\mathrm{h}$, executing $4.2\times10^{18}$ flops at $120\TFs$ aggregate
throughput (by comparison, the 2D tensor-network simulations of
Ref.~\cite{begusic2024converged} take up to one day on a single A100). In complex128,
quimb's zip-up exhausts $95\,\mathrm{GiB}$ of GPU memory at step 3 for $\chi=1024$ and
step 2 for $\chi=2048$; with a $10^{-12}$ cutoff, it completes $\chi=1024$ using
$65$--$88\GB$ at $1009\s$ per step (versus $372\s$ and $9.4\GB$ for SRC) and runs out
of memory at step 9 for $\chi=2048$.

On one GPU, the sweep sustains $17.7\TFs$ at $\chi=1024$ and $31.5\TFs$ at
$\chi=2048$ ($21\%$ and $38\%$ of a large complex64 GEMM, versus $61\TFs$ for the
whole right-to-left pass of the synthetic benchmark on one GPU). Because the gate layers have bond
dimension 2, several contractions have inner dimension 2 or 4 and are memory-bandwidth
bound; throughput rises with $\chi$ as the larger GEMMs dominate.

\begin{figure*}[t]
  \centering
  \includegraphics[width=\textwidth]{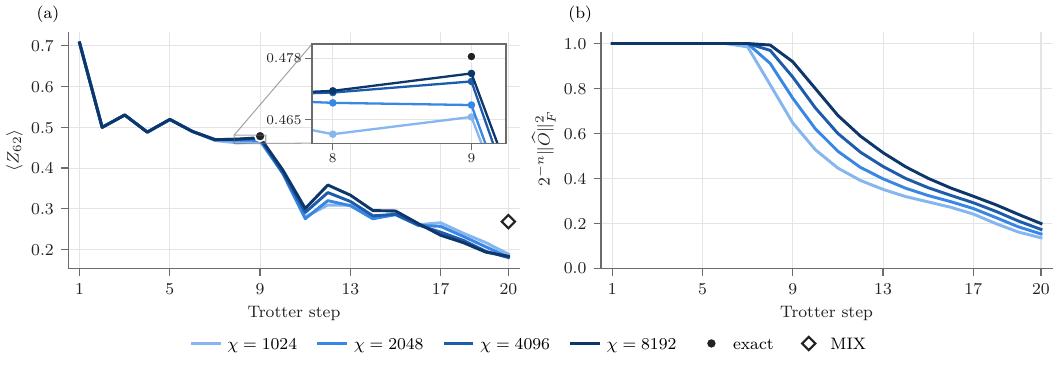}
  \caption{$\langle Z_{62}\rangle$ at $\theta_h=8\pi/32$ versus Trotter step, with SRC
  in complex64 at four bond dimensions.\ (a) Expectation value, with the
    exact 9-step value (from exact tensor-network contraction; inset: zoom around
    steps 8--9 showing monotonic convergence with $\chi$) and the 20-step MIX estimate
    of Ref.~\cite{begusic2024converged}.\ (b) Retained squared norm
    $2^{-n}\|\widehat O\|_F^2$, which equals 1 under exact evolution; in exact
    arithmetic, $1-2^{-n}\|\widehat O\|_F^2$ accumulates the normalized squared local
    compression residuals.}\label{fig:z62}
\end{figure*}

Strong scaling to eight GPUs is limited on this thin stack: each GPU sustains $8.4\TFs$
at $\chi=4096$ and $15.0\TFs$ at $\chi=8192$, and at $\chi=2048$ a one-step probe takes
$131\s$ on eight GPUs versus $106\s$ on one. At $\chi=4096$, one sub-layer spends $7\s$
in the communication-free left-to-right pass and $54\s$ in the right-to-left pass. As
discussed in \cref{sec:model}, with $\eta p=4\ell$ the rank-0 Householder QR costs
$16\eta p\ell^2=4.4\times10^{12}$ flops per site while the sketch gather, core scatter
and allgathers of $S_j$ and input cores all cross PCIe. Unlike the double-precision
A100 runs, distributed Cholesky QR cannot replace Householder QR here in complex64:
decaying singular values push the sketch condition numbers past the single-precision
threshold at every site (double-precision Gram accumulation might relax
this limit), and over PCIe exchanging $G$ full Gram matrices can already cost as much
as gathering the sketch. Here, the eight GPUs mainly pool device memory for
$\chi=8192$.

\subsection{The 20-step observable}\label{sec:app-physics}

\Cref{fig:z62} tracks $\langle Z_{62}\rangle$ and the retained squared norm
$\nu = 2^{-n}\|\widehat O\|_F^2$ across the 20 steps at $\theta_h=8\pi/32$. Under exact
unitary evolution of a Pauli operator, $\nu=1$ at every step, whereas each SRC
compression reduces $\nu$ by its squared local residual~\eqref{eq:norm-loss}. Following
Ref.~\cite{begusic2024converged}, we use $\nu$ as a convergence diagnostic: up to
single-precision roundoff, the reverse triangle inequality bounds the relative operator
error by $\|\widehat O-O(t)\|_F/\|O(t)\|_F \ge 1-\sqrt{\nu}$ (\cref{app:norm-bound}). A
small $\nu$ therefore certifies a large operator error, while $\nu\approx 1$ does not
by itself guarantee an accurate expectation value.

Up to step 7, $\nu\ge 0.98$ at all four bond dimensions and $\langle Z_{62}\rangle$
agrees within $3\times10^{-3}$. Beyond step 7, $\nu$ decays, later for larger $\chi$. At
step 9, where exact tensor-network contraction is still feasible~\cite{begusic2024converged},
the error still decreases monotonically across $\chi=1024,2048,4096,8192$
($1.3\times10^{-2}$, $1.0\times10^{-2}$,
$5.3\times10^{-3}$ and $3.6\times10^{-3}$). At step 20 it stalls: all four runs cluster
around $0.080$--$0.088$ from the MIX reference without a monotonic trend, consistent with
the non-monotonic behaviour at $\chi\le 1024$ of Anand et al.~\cite{anand2023zne}. Without $\nu$, this
agreement across an eightfold range of $\chi$ could be mistaken for convergence.
Instead, $\nu$ reaches only $0.199$ at $\chi=8192$ (up by $0.026$ from $\chi=4096$),
implying an operator error of at least $55\%$. Converging the 20-step observable thus
requires exploiting the 2D geometry or sparse Pauli
structure~\cite{tindall2024eagle,begusic2024converged}.

\Cref{tab:app-z62} in \cref{app:kicked} shows the same behaviour across four angles at
$\chi\le 2048$: errors decrease with $\chi$ at step 9, whereas at step 20 doubling
$\chi$ from $1024$ to $2048$ changes the distance to MIX by at most $0.013$. Because
$1-\sqrt{\nu}$ bounds the global operator error rather than a single matrix element, the
two can decouple: at $5\pi/32$, $\nu=0.972$ and the two bond dimensions agree to
$7\times10^{-4}$, yet both differ from MIX by $0.038$; at $10\pi/32$, SRC lies
within $0.005$ of MIX (zip-up within $0.0004$) although $\nu=0.002$.

\section{Conclusions}\label{sec:conclusions}

We have presented \code{portc}~\cite{portc}, a multi-GPU and out-of-core implementation
of successive randomized compression for stacks of tensor trains. Partitioning along the
uncontracted sketch and output-bond indices distributes both the contractions and the
stored prefix environments across GPUs and batches them across device, host and disk
memory, leaving only the QR step coupled across columns. Compressing a stack in a single
sweep pays off when a few thin gate layers dress a large state or operator, as in fused
Trotter steps, where it amortizes the orthogonalization and data movement of each
sweep; for deeper stacks or thicker layers, the $\prod_i D_i$ scaling favours
pairwise compression.

How multi-GPU and out-of-core execution pay off depends on the arithmetic intensity of
the stack and the interconnect bandwidth. When contractions dominate, distributing the
sweep reaches $3.8\times$ speed-up on four NVLink-connected GH200 GPUs ($95\%$
efficiency, predicted within $2\%$ from single-GPU measurements by our bulk-synchronous
cost model) and $5.8\times$ on eight PCIe-connected GPUs ($73\%$ efficiency), while host
and disk staging lift single-GPU memory limits. For thin stacks with cheaper contractions, collective
communication, single-precision orthogonalization and storage bandwidth eventually
become the bottleneck, though pooling device memory across GPUs still makes
very large bond dimensions accessible.

In our Heisenberg-picture simulation of the 127-qubit kicked Ising circuit, this
pooled capacity reaches operator bond dimension $\chi=8192$. Although $\langle
Z_{62}\rangle$ varies little between $\chi=1024$ and $8192$ after 20 steps, the exact
norm-loss relation reveals an operator error above $55\%$, showing both the limits of a
one-dimensional ansatz on a two-dimensional circuit and the usefulness of tracking
operator norms alongside observables. Natural technical next steps include multi-node
execution, double-precision Gram accumulation to keep distributed Cholesky QR effective
in single-precision runs, partitioning the projected environment $S_j$ across devices,
and deriving probabilistic error bounds for the adaptive Khatri--Rao
sweep~\cite{camano2025osi}.

More broadly, the structure of SRC allows large tensor-train contractions to be
distributed across devices and memory hierarchies with performance that can be
understood and predicted quantitatively. By making large-bond tensor-train products
fast and memory-bounded across multi-GPU nodes, \code{portc} expands the classical
budget available for quantum many-body dynamics and hybrid quantum--classical workflows
such as tensor-network error mitigation~\cite{filippov2023tem,filippov2024scalability},
where compressing deep products of noise-inversion and gate layers at high bond
dimension directly extends the circuit sizes and depths that can be post-processed.

\section*{Code availability}
The implementation, benchmark scripts and recorded results are available at
\url{https://github.com/Algorithmiq/portc} under the Apache-2.0
licence~\cite{portc}. The functions \code{src}, \code{apply} and \code{compress}
take and return per-site arrays, and \code{Resources} selects the devices and memory
budgets. Installation instructions, API documentation and examples are at
\url{https://algorithmiq.github.io/portc/}.

\section*{Acknowledgements}
We thank Algorithmiq's software team for helpful discussions on the accuracy of
the method. The authors acknowledge the EuroHPC Joint Undertaking for awarding
access to Leonardo, hosted by CINECA, Italy~\cite{leonardo_specs}
(\code{EHPCDEV-2026D09-033}), and JUPITER, hosted by the J\"ulich Supercomputing
Centre, Forschungszentrum J\"ulich, Germany~\cite{jupiter_specs}
(\code{EHPC-DEV-2026D11-370}).

\section*{AI usage statement}
The authors implemented and extended the method from
reference~\cite{camano2026src} and ported it to GPUs. They designed the
out-of-core and multi-GPU implementation, using Claude Opus 5.5 to refine
it. Gemini 4 Carbon assisted with manuscript editing and
mathematical review. AI was not used to generate scientific content,
references, or conclusions, and the authors checked all content for accuracy.

\hbadness=1500
\bibliography{references}
\hbadness=1000

\clearpage
\onecolumngrid
\appendix
\crefalias{section}{appendix}

\section{Notation}\label{app:notation}

\begin{table}[H]
  \caption{Notation. Byte counts use decimal units ($1\GB=10^9$
  bytes).}\label{tab:notation}
  \small
  \begin{tabular*}{\textwidth}{@{\extracolsep{\fill}}ll@{}}
    \toprule
    Symbol                     & Meaning                                                                                    \\
    \midrule
    $n$, $k$                   & number of sites, number of layers in the stack                                             \\
    $\Pi_{D,j}$                & uncompressed bond dimension of the exact product to the left of site $j$                   \\
    $D_i$, $\Pi_D$, $\Sigma_D$ & bulk bond dimension of layer $i$, $\prod_{i}D_i$, $\sum_{i}D_i$                            \\
    $d$, $p_j=\dim x_0\dim x_k$& physical leg dimension, open physical dimension of site $j$                                \\
    $\chi$                     & target (maximum) output bond dimension                                                     \\
    $\ell\ge\chi$              & sketch size; $\ell=\chi$ without oversampling                                              \\
    $r_j$, $\eta_j=r_{j+1}$    & output bond to the left of site $j$, rows of the projected environment entering it         \\
    $C_j$, $S_j$               & sketched prefix environment ($\ell\times \Pi_{D,j}$), projected suffix environment ($r_j\times \Pi_{D,j}$) \\
    $G$; $P_g$, $R_g$, $T_g$   & number of GPUs; blocks of sketch indices, output bonds and rows of $M_j$ owned by rank $g$ \\
    $b$                        & batch size along the free index of a contraction                                           \\
    $e$                        & bytes per element ($16$ for complex128)                                                    \\
    $K_j$                      & bytes of the cores of site $j$, over all layers                                            \\
    \bottomrule
  \end{tabular*}
\end{table}

\section{Contraction-path diagnostics}\label{app:paths}

The initial A100 runs used the default heuristic of opt\_einsum. For the synthetic
benchmark, its projection path contracts $S_{j+1}$ with the input cores before the
batch of output rows. For batches of at least 32 rows, this creates an intermediate of
$\eta\Pi_D p=8.2\times10^9$ elements ($131\GB$ in complex128), with a planned peak of
$265$--$283\GB$. The resulting memory constraint forces batches of 16 rows, for which
the heuristic path still performs approximately twice the work of the
dynamic-programming path: on one A100, the projection takes $22.1$ versus
$11.0\,\mathrm{ms}$ per row at this batch size. Dynamic programming allows batches of
32--128 rows with planned peaks of 10--25\,GB\@.

\Cref{tab:paths} compares complete sweeps at $D_3=1000$ and $\chi=500$. For these
shapes, the heuristic sketch and projection paths require $3.9$ and $2.0$ times the
optimal flop counts, respectively. The gain is in the right-to-left pass; the
first-pass times change little. The outputs agree to a relative Frobenius distance
below $4\times10^{-13}$, and the path search adds $0.03$--$0.10\s$ per sweep.

\begin{table}[H]
  \caption{Path-selection comparison on A100 GPUs, complex128, $D_3=1000$, $\chi=500$;
  median of three warm runs with the same planned batches.}\label{tab:paths}
  \small
  \begin{tabular*}{\textwidth}{@{\extracolsep{\fill}}rrrrr@{}}
    \toprule
    GPUs & Heuristic (s) & Dynamic programming (s) & Speed-up & Right-to-left (s), heuristic / DP \\
    \midrule
    1    & 109.7         & 56.2                    & 1.95     & 95.4 / 42.3                       \\
    4    & 32.4          & 17.4                    & 1.86     & 28.6 / 13.6                       \\
    \bottomrule
  \end{tabular*}
\end{table}

\section{Supplementary benchmarks}\label{app:benchmarks}

\Cref{tab:depth} compares a single sweep over a stack of depth $k$ against pairwise
application for the experiments of \cref{sec:exp-stacks}.

\begin{table}[H]
  \caption{One sweep over a stack of depth $k$ ($k-1$ MPO layers and an MPS) versus
    pairwise application: 30 sites, MPS bond 64, $\chi=64$, best of three on a laptop
    CPU (AMD Ryzen 7 7840U).}\label{tab:depth}
  \small
  \begin{tabular*}{\textwidth}{@{\extracolsep{\fill}}lrrrr@{}}
    \toprule
    MPO layers     & $k$ & One sweep (s) & Pairwise (s) & Ratio \\
    \midrule
    Trotter        & 2   & 0.0764        & 0.0642       & 1.19  \\
    Trotter        & 3   & 0.0936        & 0.147        & 0.64  \\
    Trotter        & 4   & 0.201         & 0.207        & 0.98  \\
    Trotter        & 5   & 0.384         & 0.283        & 1.36  \\
    random, $D=8$  & 3   & 1.67          & 0.52         & 3.21  \\
    random, $D=8$  & 4   & 24.7          & 0.837        & 29.5  \\
    random, $D=16$ & 4   & 316           & 2.33         & 136   \\
    \bottomrule
  \end{tabular*}
\end{table}

\begin{figure}[H]
  \centering
  \includegraphics[width=\textwidth]{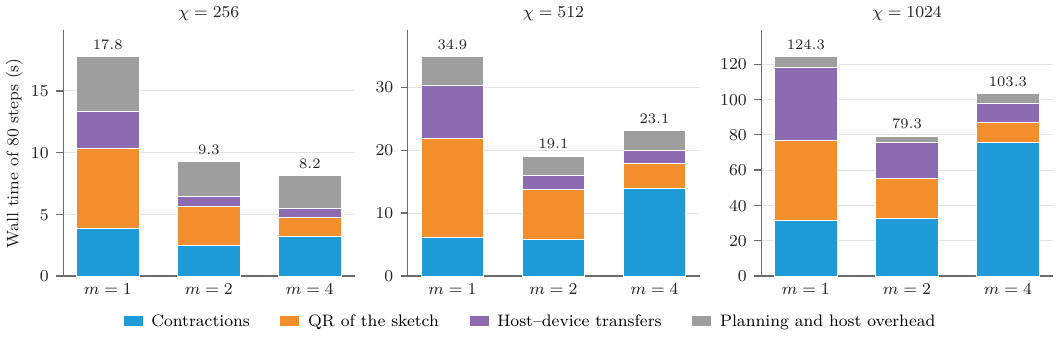}
  \caption{Wall time of the 80-step quench of 50 qubits on one A100, complex128,
    dynamic-programming paths, fusing $m$ Trotter steps per sweep, split into
    contractions, the Householder QR of the sketch, host--device transfers of the
    state, and planning and host overhead. Totals are instrumented wall times.}%
  \label{fig:quench-breakdown}
\end{figure}

In the 50-qubit quench (\cref{sec:exp-quench}), doubling the number of steps per
sweep halves the total QR time, but fusing four steps doubles the contraction work
(\cref{fig:quench-breakdown}). The leading contraction cost per sweep scales as
$2^m\chi^3$, and the number of sweeps falls as $1/m$. The total contraction flop
count is therefore the same for $m=1$ and $m=2$, and twice as large for $m=4$.
Counting the inputs and outputs of each pairwise contraction gives arithmetic
intensities of $19$--$170\,\mathrm{flop/B}$, above the ridge point of
$15\,\mathrm{flop/B}$ set by the measured GEMM rate and the
$1.41\,\mathrm{TB/s}$ device-to-device copy bandwidth. The QR also has high
arithmetic intensity ($127$--$508\,\mathrm{flop/B}$), but its speed is limited by
panel factorizations and host synchronization rather than by bandwidth. We time
the contractions and QR with CUDA events, count flops from the executed paths,
and attribute the remaining SRC call time to host--device state transfers, path
planning and Python overhead. Instrumentation changes wall times by $-9\%$ to
$+10\%$, within the run-to-run spread.

The contraction path with the fewest flops is not always the fastest. Device
counters at one bulk site at $\chi=512$ show that the path for $m=4$ inserts more
tensor transposes than the path for $m=1$: the transposes take $21\%$ of the site's
kernel time rather than $12\%$, and the measured arithmetic intensity falls from
$50$ to $19\,\mathrm{flop/B}$. At $\chi=512$ and $m=4$, opt\_einsum's default
heuristic uses $1.5$ times as many flops for the sketches and projections as
dynamic programming, but sustains $16\TFs$ instead of $10\TFs$. The full evolution
thus takes $19.1\s$ rather than $23.0\s$. For the large-bond stacks of
\cref{app:paths}, by contrast, dynamic programming halves the sweep time.

\section{The distributed Cholesky QR}\label{app:cholqr}

\Cref{alg:cholqr} lists the three-pass shifted Cholesky QR of \cref{sec:dist-qr} as executed by
rank $g$ on its rows $X=M_j[T_g,:]$ of the sketch after the first all-to-all. Each pass
exchanges one $\ell\times\ell$ Gram matrix; everything else is local or replicated. The
partial Gram matrices are summed in rank order, so that every rank holds
bitwise-identical $W$, takes the same branches and declines at the same pass. The shift
follows Ref.~\cite{fukaya2020shifted}, with $\operatorname{tr}W\ge\|M_j\|_2^2$ in place
of the norm. A pass whose input Gram matrix lies within $0.1$ of the identity is the
last one; if the input of the third pass is still not within $0.1$, the sketch is
declined. The sweep calls \cref{alg:cholqr} at site $j$ only if $G>1$ and
$\eta_{j}p_j\ge\ell\ge1024$, unless a backoff after earlier declines has suspended it;
when it declines or is not called, the site uses the Householder QR on rank 0.

\begin{algorithm}[H]
  \caption{Three-pass shifted Cholesky QR of $M_j\in\C^{m\times\ell}$, $m=\eta_{j}p_j$, as executed
    by rank $g$. $X^{(h)}$ denotes the current rows $T_h$ held by rank $h$, and $u$ is
    the unit roundoff.}\label{alg:cholqr}
  \begin{algorithmic}[1]
    \REQUIRE rows $X=M_j[T_g,:]$; relative cutoff $\tau\ge0$
    \ENSURE rows $Q_j[T_g,:]$ of an orthonormal basis for the range of $M_j$, optionally
    truncated, or \textsc{declined} on every rank
    \STATE $q\gets0$; $\mathit{done}\gets$ \textbf{false}
    \WHILE{\textbf{not} $\mathit{done}$}
    \STATE $q\gets q+1$
    \STATE $W\gets\sum_{h=0}^{G-1}{X^{(h)}}^\dagger X^{(h)}$, summed in rank order
    \hfill\COMMENT{\textsc{Allgather} of $X^\dagger X$}
    \STATE $W\gets(W+W^\dagger)/2$
    \IF{$q=1$}
    \STATE $W\gets W+11\,\bigl(m\ell+\ell(\ell+1)\bigr)\,u\,\operatorname{tr}(W)\,I$
    \hfill\COMMENT{shifted pass}
    \ELSIF{$\|W-I\|_F\le0.1$}
    \STATE $\mathit{done}\gets$ \textbf{true} \hfill\COMMENT{last pass}
    \ELSIF{$q=3$}
    \RETURN \textsc{declined}
    \ENDIF
    \STATE $L_q\gets$ Cholesky factor of $W$, $W=L_{q}L_q^\dagger$
    \IF{the factorization fails or $\min_i{\operatorname{Re}(L_q)}_{ii}\le0$}
    \RETURN \textsc{declined}
    \ENDIF
    \STATE $X\gets X L_q^{-\dagger}$ \hfill\COMMENT{triangular solve, local}
    \ENDWHILE
    \IF{$\tau>0$}
    \STATE $R_j\gets L_q^\dagger\cdots L_1^\dagger$ \hfill\COMMENT{$M_j=Q_{j}R_j$, on
    every rank}
    \STATE $U\Sigma V^\dagger\gets$ SVD of $R_j$
    \STATE $r\gets\max\bigl(1,\,\#\{i:\sigma_i\ge\tau\sigma_1\}\bigr)$
    \STATE $X\gets X\,U_{[:,1:r]}$
    \ENDIF
    \RETURN $X$
  \end{algorithmic}
\end{algorithm}

\section{Memory of the sweep}\label{app:memory}

\Cref{tab:memory} lists the objects that the distributed sweep of \cref{sec:ooc}
keeps in memory, with their sizes for the synthetic benchmark.

\begin{table}[H]
  \caption{Memory of the sweep for the synthetic benchmark ($G=4$).
    The sizes are not additive since object lifetimes determine the simultaneous
    residency.}\label{tab:memory}
  \small
  \begin{tabular*}{\textwidth}{@{\extracolsep{\fill}}lllr@{}}
    \toprule
    Object                             & Location & Size                                     & Benchmark       \\
    \midrule
    environments $C_j[P_g]$, all sites & tiered   & $|P_g|\,e\sum_{j=2}^n \Pi_{D,j}$ per rank& $401/G\GB$      \\
    cores of a site, prefetched        & device   & $(1+\text{prefetch})\,K$                 & $8.2\GB$        \\
    projected environment $S_{j+1}$    & device   & $\eta\Pi_D e$                            & $8.2\GB$        \\
    new $S_j$ and local rows           & device   & up to $(r_j+|R_g|)\,\Pi_{D,j} e$         & $10.2\GB$       \\
    rows of $M_j$ and of $Q_j$         & device   & $\approx 2\,\eta p \ell e/G$             & $0.5\GB$        \\
    batch working set                  & device   & $\propto b\,\Pi_D\,e$                    & planned         \\
    output cores                       & host     & $e\sum_j r_j\eta_{j}p_j$                 & $\lesssim50\GB$ \\
    \bottomrule
  \end{tabular*}
\end{table}

\section{Kicked Ising runs}\label{app:kicked}

\Cref{tab:app-z62} lists the observable errors and retained operator norms across the
four angles of \cref{sec:app-physics}.

\begin{table}[H]
  \caption{$\langle Z_{62}\rangle$ with SRC at $\chi=1024$ and $2048$ (complex64,
    cutoff 0): absolute error after 9 steps against the exact values and after 20
    steps against the MIX estimates of Ref.~\cite{begusic2024converged},
    zip-up at $\chi=1024$ (complex128, cutoff $10^{-12}$), and the
    retained squared norm at $\chi=2048$ after 20 steps.}\label{tab:app-z62}
  \small
  \begin{tabular*}{\textwidth}{@{\extracolsep{\fill}}lrrrrrrr@{}}
    \toprule
               &       & \multicolumn{2}{c}{9 steps, error} & \multicolumn{3}{c}{20 steps, distance to MIX} & 20 steps                                                    \\
    \cmidrule(lr){3-4}\cmidrule(lr){5-7}
    $\theta_h$ & MIX   & 1024                               & 2048                                          & 1024     & 2048  & zip-up 1024 & $2^{-n}\|\widehat O\|_F^2$ \\
    \midrule
    $5\pi/32$  & 0.797 & $4.6\times10^{-4}$                 & $1.3\times10^{-5}$                            & 0.038    & 0.038 & 0.049       & 0.972                      \\
    $7\pi/32$  & 0.511 & $1.3\times10^{-3}$                 & $1.4\times10^{-4}$                            & 0.182    & 0.195 & 0.200       & 0.477                      \\
    $8\pi/32$  & 0.269 & $1.3\times10^{-2}$                 & $1.0\times10^{-2}$                            & 0.080    & 0.087 & 0.114       & 0.153                      \\
    $10\pi/32$ & 0.016 & $3.0\times10^{-2}$                 & $2.3\times10^{-2}$                            & 0.005    & 0.003 & 0.0004      & 0.002                      \\
    \bottomrule
  \end{tabular*}
\end{table}

\section{A norm-loss bound on the operator error}\label{app:norm-bound}

Let $O(t)=V^\dagger OV$ be the exactly evolved observable, where $V$ is the unitary of
the gate layers applied so far, and let $\widehat O$ be the MPO returned by the
compressed evolution. We measure the relative operator error and the retained squared
norm,
\begin{equation}
  \varepsilon=\frac{\|\widehat O-O(t)\|_F}{\|O(t)\|_F},
  \qquad
  \nu=2^{-n}\|\widehat O\|_F^2 .
\end{equation}
Conjugation by a unitary preserves the Frobenius norm,
$\|O(t)\|_F^2=\operatorname{tr}(V^\dagger O^\dagger VV^\dagger OV)
=\operatorname{tr}(O^\dagger O)=\|O\|_F^2$. For a Pauli string, $O^\dagger O=I$, so
$\|O(t)\|_F^2=\operatorname{tr}I=2^n$ at every step, and $\nu=1$ under exact evolution.
Expanding the error and applying the Cauchy--Schwarz inequality
$\operatorname{Re}\langle O(t),\widehat O\rangle\le\|O(t)\|_F\|\widehat O\|_F$, with
$\langle A,B\rangle=\operatorname{tr}(A^\dagger B)$, gives
\begin{equation}\label{eq:norm-bound}
  \varepsilon^2
  =1+\nu-\frac{2\operatorname{Re}\langle O(t),\widehat O\rangle}{2^n}
  \ge1+\nu-2\sqrt\nu=\bigl(1-\sqrt\nu\bigr)^2 ,
\end{equation}
that is, $\varepsilon\ge|1-\sqrt\nu|$, the reverse triangle inequality. In exact
arithmetic, $\nu\le1$: each compression removes the norm of its residual
by~\eqref{eq:norm-loss}, the truncation sweep only discards Schmidt values, and the
gate layers preserve the norm. Hence
\begin{equation}
  \varepsilon\ge1-\sqrt\nu .
\end{equation}
Equality holds when $\widehat O$ is a nonnegative multiple of $O(t)$, so the bound is
tight when compression mainly shrinks the operator and loose when it also rotates it.
Conversely,~\eqref{eq:norm-bound} with $\nu=1$ permits any $\varepsilon\le2$, so a
retained norm near unity does not by itself certify an accurate operator or expectation
value. At $\chi=8192$ after 20 steps, $\nu=0.199$ gives
$\varepsilon\ge1-\sqrt{0.199}=0.554$, the bound quoted in \cref{sec:app-physics}.

\end{document}

%% file: figures/tikz/styles.tex
\usetikzlibrary{arrows.meta,positioning,calc,shapes.geometric,patterns,decorations.pathreplacing,fit,backgrounds}

\newcommand{\dsmall}{\fontsize{9}{11}\selectfont}
\newcommand{\dscript}{\fontsize{7}{8}\selectfont}
\newcommand{\dtiny}{\fontsize{5}{6}\selectfont}

\definecolor{srcSketch}{HTML}{F2C12E}   
\definecolor{srcLayer}{HTML}{1A9E77}    
\definecolor{srcState}{HTML}{1F9BD7}    
\definecolor{srcEnv}{HTML}{9E9E9E}      
\definecolor{srcProj}{HTML}{E58C8A}     
\definecolor{srcIso}{HTML}{B5B5B5}      
\definecolor{srcRoot}{HTML}{F28E2B}     
\definecolor{srcComm}{HTML}{59A14F}     
\definecolor{srcIO}{HTML}{8C6BB1}       
\definecolor{srcInk}{HTML}{222222}

\tikzset{
  tensor/.style={circle, draw=srcInk, line width=0.6pt, minimum size=4.2mm, inner sep=0pt},
  layer/.style={tensor, fill=srcLayer},
  state/.style={tensor, fill=srcState},
  sketch/.style={tensor, fill=srcSketch},
  env/.style={rectangle, rounded corners=1pt, draw=srcInk, line width=0.6pt, fill=srcEnv, minimum width=4.2mm},
  proj/.style={env, fill=srcProj},
  iso/.style={isosceles triangle, isosceles triangle apex angle=60, draw=srcInk, line width=0.6pt,
    fill=srcIso, minimum size=6mm, inner sep=0pt},
  hyper/.style={circle, fill=srcInk, minimum size=1.6mm, inner sep=0pt},
  bond/.style={line width=0.8pt, draw=srcInk},
  leg/.style={bond},
  idx/.style={font=\dscript, inner sep=1pt},
  panel/.style={font=\dsmall\bfseries, anchor=north west},
  op/.style={-{Stealth[length=2mm]}, line width=0.7pt, draw=srcInk},
  task/.style={draw=srcInk, line width=0.4pt, minimum height=4.6mm, inner sep=1pt, font=\dscript, anchor=west},
  idle/.style={draw=srcInk!50, line width=0.4pt, pattern=north east lines, pattern color=srcInk!35,
    minimum height=4.6mm, anchor=west},
  flow/.style={-{Stealth[length=1.8mm]}, line width=0.7pt},
  box/.style={draw=srcInk, line width=0.6pt, rounded corners=2pt, align=center, font=\dscript},
}

%% file: figures/tikz/src_steps.tex
\begin{tikzpicture}[x=1cm, y=1cm]
  \tikzset{
    pics/stack/.style={code={
      \node[layer] (-1) at (0, 0) {};
      \node[layer] (-2) at (0,-0.8) {};
      \node[state] (-3) at (0,-1.6) {};
      \draw[bond] (-1) -- (-2) -- (-3);
    }},
    tall/.style={minimum height=21mm},
  }
  \newcommand{\envat}[3]{\node[#1, tall] (#2) at (#3,-0.8) {};}
  \newcommand{\bondsleft}[2]{\foreach \i/\y in {1/0,2/-0.8,3/-1.6} \draw[bond] (#1,\y) -- (#2-\i);}
  \newcommand{\bondsright}[2]{\foreach \i/\y in {1/0,2/-0.8,3/-1.6} \draw[bond] (#2-\i) -- (#1,\y);}
  \newcommand{\isometry}[4]{%
    \draw[line width=0.6pt, draw=srcInk, fill=srcIso, line join=round]
      ($(#1,#3)+(-0.18,0)$) -- ($(#2,#3)+(0.18,0)$) -- ($({(#1+#2)/2},#3+0.55)$) -- cycle;
    \coordinate (#4) at ($({(#1+#2)/2},#3+0.55)$);}

  \begin{scope}[xshift=-3.90cm]
    \node[panel] at (-0.75,2.15) {(a)};
    \begin{scope}[on background layer]
      \filldraw[fill=srcInk!6, draw=srcInk!30, line width=0.45pt, rounded corners=2pt]
        (0.71,0.31) rectangle (1.29,-1.91);
    \end{scope}
    \node[idx, anchor=east] at (-0.05, 0) {$A^{(1)}$};
    \node[idx, anchor=east] at (-0.05,-0.8) {$A^{(2)}$};
    \node[idx, anchor=east] at (-0.05,-1.6) {$\psi$};
    \pic (s1) at (0.30, 0) {stack};
    \pic (sj) at (1.00, 0) {stack};
    \pic (sk) at (1.70, 0) {stack};
    \pic (sn) at (2.82, 0) {stack};
    \foreach \s in {s1, sj, sk, sn} \draw[leg] (\s-1) -- ++(0,0.55);
    \node[idx, anchor=south] at (1.00,0.55) {$x_0$};
    \foreach \i in {1,2,3} {
      \draw[bond] (s1-\i) -- (sj-\i) -- (sk-\i);
      \draw[bond] (sk-\i.east) -- ++(0.18,0);
      \draw[bond] (sn-\i.west) -- ++(-0.18,0);
      \node[font=\dscript] at ($(sk-\i)!0.5!(sn-\i)$) {$\cdots$};
    }
    \node[idx, anchor=north] at (0.30,-1.96) {$1$};
    \node[idx, anchor=north] at (1.00,-1.96) {$j$};
    \node[idx, anchor=north] at (2.82,-1.96) {$n$};
  \end{scope}

  \node[panel] at (-0.45,2.15) {(b)};
  \envat{env}{Ca}{0}
  \node[idx, below=1pt of Ca] {$C_j$};
  \pic (a) at (1.1,0) {stack};
  \bondsleft{0.21}{a}
  \bondsright{1.8}{a}
  \node[idx, anchor=west] at (1.8,-0.8) {$\beta$};
  \node[sketch] (Oa) at (1.1,0.8) {};
  \node[idx, anchor=west] at (1.32,0.8) {$\Omega_j$};
  \draw[leg] (a-1) -- (Oa);
  \node[hyper] (ha) at (1.1,1.5) {};
  \draw[leg] (0,0.25) -- (0,1.5) -- (ha) -- (Oa);
  \draw[leg] (ha) -- (1.8,1.5) node[idx, anchor=west] {$s$};
  \node[idx, anchor=east] at (1.0,-0.4) {$x_1$};
  \node[idx, anchor=east] at (1.0,-1.2) {$x_2$};
  \node[idx, anchor=east] at (1.0,0.42) {$x_0$};
  \node at (2.35,-0.8) {$=$};
  \envat{env}{Cb}{2.85}
  \node[idx, below=1pt of Cb] {$C_{j+1}$};
  \draw[leg] (2.85,0.25) -- (2.85,1.5) node[idx, anchor=south] {$s$};
  \foreach \y in {0,-0.8,-1.6} \draw[bond] (3.06,\y) -- (3.4,\y);

  \begin{scope}[xshift=4.5cm]
    \node[panel] at (-0.45,2.15) {(c)};
    \envat{env}{Cc}{0}
    \node[idx, below=1pt of Cc] {$C_j$};
    \pic (b) at (1.0,0) {stack};
    \bondsleft{0.21}{b}
    \envat{proj}{Sb}{2.0}
    \node[idx, below=1pt of Sb] {$S_{j+1}$};
    \bondsright{1.79}{b}
    \draw[leg] (0,0.25) -- (0,1.5) node[idx, anchor=south] {$s$};
    \draw[leg] (b-1) -- (1.0,1.5) node[idx, anchor=south] {$x_0$};
    \draw[leg] (2.0,0.25) -- (2.0,1.5) node[idx, anchor=south] {$\eta$};
  \end{scope}

  \begin{scope}[xshift=7.9cm]
    \node[panel] at (-0.45,2.15) {(d)};
    \node[env, fill=white, minimum width=12mm, minimum height=5mm] (M) at (0.5,0.75) {\dscript $M_j$};
    \draw[leg] (M.north) -- ++(0,0.5) node[idx, anchor=south] {$s$};
    \draw[leg] ([xshift=-3mm]M.south) -- ++(0,-0.3) node[idx, anchor=north] {$x_0$};
    \draw[leg] ([xshift=3mm]M.south) -- ++(0,-0.3) node[idx, anchor=north] {$\eta$};
    \draw[op] (0.5,-0.05) -- (0.5,-0.7) node[midway, right, idx] {QR};
    \isometry{0.2}{0.8}{-1.65}{Ga}
    \node[idx] at (0.5,-1.43) {$\Gamma_j$};
    \draw[leg] (0.2,-1.65) -- (0.2,-1.95) node[idx, anchor=north] {$x_0$};
    \draw[leg] (0.8,-1.65) -- (0.8,-1.95) node[idx, anchor=north] {$\eta$};
    \draw[leg] (Ga) -- ++(0,0.25) node[idx, anchor=west] {$\rho$};
  \end{scope}

  \begin{scope}[xshift=10.0cm]
    \node[panel] at (-0.45,2.15) {(e)};
    \pic (d) at (0.6,0) {stack};
    \foreach \i/\y in {1/0,2/-0.8,3/-1.6} \draw[bond] (-0.05,\y) -- (d-\i);
    \node[idx, anchor=east] at (-0.05,-0.8) {$\alpha$};
    \envat{proj}{Sd}{1.6}
    \node[idx, below=1pt of Sd] {$S_{j+1}$};
    \bondsright{1.39}{d}
    \isometry{0.6}{1.6}{0.65}{Gd}
    \node[idx] at (1.1,0.85) {$\overline{\Gamma}_j$};
    \draw[leg] (d-1) -- (0.6,0.65);
    \draw[leg] (1.6,0.25) -- (1.6,0.65);
    \draw[leg] (Gd) -- ++(0,0.3) node[idx, anchor=south] {$\rho$};
    \node at (2.25,-0.8) {$=$};
    \envat{proj}{Sn}{3.05}
    \node[idx, below=1pt of Sn] {$S_j$};
    \draw[leg] (3.05,0.25) -- (3.05,1.5) node[idx, anchor=south] {$\rho$};
    \foreach \y in {0,-0.8,-1.6} \draw[bond] (2.84,\y) -- (2.55,\y);
  \end{scope}
\end{tikzpicture}

%% file: figures/tikz/distributed_site.tex
\begin{tikzpicture}[x=1cm, y=1cm]
  \colorlet{rank0}{srcState!12}
  \colorlet{rank1}{srcState!30}
  \colorlet{rank2}{srcState!48}
  \colorlet{rank3}{srcState!66}
  \tikzset{blk/.style={draw=srcInk, line width=0.5pt}}
  \newcommand{\rowblocks}[5]{%
    \foreach \g in {0,1,2,3}
      \fill[rank\g] ($(#1,#2)+(0,-\g*#5/4)$) rectangle ++(#4,-#5/4);
    \foreach \g in {1,2,3}
      \draw[line width=0.3pt, srcInk!60] ($(#1,#2)+(0,-\g*#5/4)$) -- ++(#4,0);
    \draw[blk] (#1,#2) rectangle ++(#4,-#5);
    \coordinate (#3) at ($(#1,#2)+(#4/2,0)$);}

  \node[panel] at (-0.75,1.15) {(a)};
  \rowblocks{0}{0.6}{C}{1.9}{1.6}
  \node[idx, above=1pt of C] {$C_j$, $\ell\times \Pi_{D,j}$};
  \foreach \g in {0,1,2,3}
    \node[idx, anchor=east] at (0,0.6-0.2-\g*0.4) {$P_{\g}$};
  \fill[srcProj!60] (2.45,0.6) rectangle ++(1.3,-1.6);
  \draw[blk] (2.45,0.6) rectangle ++(1.3,-1.6);
  \node[idx, above=1pt] at (3.1,0.6) {$S_{j+1}$, $\eta\times \Pi_{D,j+1}$};
  \node[idx, align=center] at (3.1,-0.2) {all\\ranks};
  \rowblocks{0}{-1.55}{G}{1.9}{0.8}
  \node[idx, below=1pt] at (0.95,-2.35) {$\Gamma_j$, $r_j\times\eta_jp_j$};
  \foreach \g in {0,1,2,3}
    \node[idx, anchor=east] at (0,-1.55-0.1-\g*0.2) {$R_{\g}$};
  \rowblocks{2.45}{-1.55}{M}{1.3}{0.8}
  \node[idx, below=1pt] at (3.1,-2.35) {$M_j$, $\eta_jp_j\times\ell$};
  \foreach \g in {0,1,2,3}
    \node[idx, anchor=west] at (3.75,-1.55-0.1-\g*0.2) {$T_{\g}$};

  \begin{scope}[xshift=5.4cm]
    \node[panel] at (-1.25,1.15) {(b)};
    \def\h{0.23}
    \newcommand{\task}[5]{%
      \draw[line width=0.4pt, draw=srcInk, fill=#4] (#2,#1-\h) rectangle (#3,#1+\h);
      \node[font=\dscript] at ({(#2+#3)/2},#1) {#5};}
    \newcommand{\idle}[3]{%
      \draw[line width=0.3pt, draw=srcInk!40, pattern=north east lines, pattern color=srcInk!35]
        (#2,#1-\h) rectangle (#3,#1+\h);}
    \newcommand{\io}[3]{\fill[srcIO] (#2,#1-\h-0.13) rectangle (#3,#1-\h-0.05);}

    \foreach \g/\se/\pe in {0/3.15/7.8, 1/3.3/7.9, 2/3.05/7.75, 3/3.25/7.95} {
      \pgfmathsetmacro{\y}{0.3-0.75*\g}
      \node[idx, anchor=east] at (-0.15,\y) {rank \g};
      \fill[rank\g] (-0.12,\y-0.08) rectangle ++(0.1,0.16);
      \io{\y}{0}{0.35} \io{\y}{0.5}{0.85} \io{\y}{1.3}{1.65} \io{\y}{2.1}{2.45}
      \idle{\y}{0}{0.35}
      \task{\y}{0.35}{\se}{srcState!35}{sketch $Y_j[\cdot,P_\g]$}
      \idle{\y}{\se}{3.4}
      \task{\y}{3.4}{3.65}{srcComm!70}{}
      \task{\y}{3.65}{4.75}{srcRoot!70}{}
      \foreach \x in {3.88, 4.3}
        \draw[line width=0.3pt, draw=srcInk, fill=srcComm!70] (\x,\y-\h) rectangle ++(0.07,2*\h);
      \task{\y}{4.75}{5.0}{srcComm!70}{}
      \io{\y}{5.0}{5.3}
      \task{\y}{5.0}{\pe}{srcState!35}{project $S_j[R_\g,:]$}
      \idle{\y}{\pe}{8.05}
      \task{\y}{8.05}{8.5}{srcComm!70}{}
      \io{\y}{8.5}{8.85}
      \idle{\y}{8.5}{8.85}
      \draw[line width=0.4pt, draw=srcInk!50, fill=srcState!15] (8.85,\y-\h) rectangle (9.4,\y+\h);
    }
    \node[idx, anchor=south east] at (3.65,0.55) {all-to-all};
    \node[idx, anchor=south west] at (4.75,0.55) {all-to-all};
    \node[idx, anchor=south] at (8.275,0.55) {allgather};
    \draw[decorate, decoration={brace, amplitude=3pt}, line width=0.5pt]
      (3.67,0.84) -- (4.73,0.84) node[midway, above=3pt, idx] {Cholesky QR};
    \draw[decorate, decoration={brace, amplitude=3pt, mirror}, line width=0.5pt]
      (0,-2.4) -- (8.5,-2.4) node[midway, below=3pt, idx] {site $j$};
    \draw[decorate, decoration={brace, amplitude=3pt, mirror}, line width=0.5pt]
      (8.52,-2.4) -- (9.4,-2.4) node[midway, below=3pt, idx] {site $j-1$};
    \draw[op] (0,-2.95) -- (9.4,-2.95) node[idx, anchor=north east] at (9.4,-3.0) {time};
    \foreach \i/\c/\t in {0/srcState!35/local contraction, 1/srcRoot!70/Cholesky QR, 2/srcComm!70/collective} {
      \fill[\c, draw=srcInk, line width=0.3pt] (\i*2.6,-3.45) rectangle ++(0.3,-0.18);
      \node[idx, anchor=west] at (0.33+\i*2.6,-3.54) {\t};
    }
    \draw[line width=0.3pt, draw=srcInk!40, pattern=north east lines, pattern color=srcInk!35]
      (7.8,-3.45) rectangle ++(0.3,-0.18);
    \node[idx, anchor=west] at (8.13,-3.54) {idle};
    \fill[srcIO] (0,-3.77) rectangle ++(0.3,-0.08);
    \node[idx, anchor=west] at (0.33,-3.81) {transfer: a batch of $C_j$ read (the first one is exposed), rows of $\Gamma_j$ to the host};
  \end{scope}
\end{tikzpicture}

%% file: figures/tikz/ooc_pipeline.tex
\begin{tikzpicture}[x=1cm, y=1cm]
  \tikzset{
    tier/.style={draw=srcInk, line width=0.6pt, rounded corners=3pt, fill=#1},
    item/.style={box, fill=white, minimum height=6mm, text width=#1},
    item/.default=17mm,
    ring/.style={item=#1, fill=srcIO!12},
    ring/.default=17mm,
    lbl/.style={font=\dscript\itshape, inner sep=1.5pt, align=center},
    bw/.style={font=\dtiny, inner sep=1pt, text=srcInk!75},
    head/.style={font=\dscript\bfseries, anchor=north, align=center},
  }
  \def\rc{0.35} \def\rw{-0.9} \def\rr{-2.1} \def\rbot{-3.85}

  \draw[tier=srcEnv!12] (0,1.25) rectangle (2.2,\rbot);
  \node[head] at (1.1,1.2) {file system};
  \node[bw, anchor=south] at (1.1,\rbot+0.05) {GB/s, shared};
  \draw[tier=srcSketch!12] (2.85,1.25) rectangle (7.55,\rbot);
  \node[head] at (5.2,1.2) {host memory};
  \node[bw, anchor=south] at (5.2,\rbot+0.05) {DDR4, $\sim\!200$\,GB/s};
  \draw[tier=srcLayer!10] (9.25,1.25) rectangle (11.75,\rbot);
  \node[head] at (10.5,1.2) {GPU $g$};
  \node[bw, anchor=south] at (10.5,\rbot+0.05) {HBM2e, $\sim\!2$\,TB/s};
  \draw[tier=srcLayer!10] (13.35,1.25) rectangle (14.4,-0.45);
  \node[font=\dscript\bfseries, align=center] at (13.875,0.4) {other\\GPUs};

  \node[item] (input) at (1.1,\rc) {input stack\\on disk};
  \node[item, minimum height=16mm] (scratch) at (1.1,-1.5) {scratch tier\\$C_j$, one file\\per site};
  \node[ring] (cring) at (6.4,\rc) {core ring\\(pinned)};
  \node[item=10mm, minimum height=16mm] (htier) at (3.65,-1.5) {host\\tier\\$C_j$};
  \node[ring] (wring) at (6.4,\rw) {write ring\\(pinned)};
  \node[ring] (rring) at (6.4,\rr) {read ring\\(pinned)};
  \node[item=19mm] (cores) at (10.5,\rc) {cores of site $j$};
  \node[item=19mm, minimum height=16mm, fill=srcState!15] (kern) at (10.5,-1.5) {kernels,\\one batch\\at a time};
  \node[item=19mm, fill=srcEnv!25] (dtier) at (10.5,-3.05) {device tier $C_j$};

  \draw[flow, srcInk] (input) -- (cring) node[midway, above=1.5pt, lbl] {loader};
  \draw[flow, srcInk] (cring) -- (cores)
    node[midway, above=1.5pt, lbl] {$1/G$ slice}
    node[midway, below=1.5pt, bw, align=center] {PCIe 4.0\\32\,GB/s};
  \draw[{Stealth[length=1.8mm]}-{Stealth[length=1.8mm]}, line width=0.7pt, srcComm!80!black]
    (cores.east) -- (13.35,\rc)
    node[pos=0.56, above=1.5pt, lbl] {allgather}
    node[pos=0.56, below=1.5pt, bw, align=center] {NVLink\\100\,GB/s};
  \draw[flow, srcInk] (cores) -- (kern);

  \draw[flow, srcIO] (kern.west |- 0,\rw) -- (wring.east)
    node[midway, above=1.5pt, lbl] {copy stream};
  \draw[flow, srcIO] (wring.west) -- (htier.east |- 0,\rw)
    node[midway, above=2pt, lbl] {writer};
  \draw[flow, srcIO] ([xshift=-2mm]wring.north) to[out=152, in=25]
    node[pos=0.52, above=1pt, lbl] {or} (scratch.north east);

  \draw[flow, srcIO] (htier.east |- 0,\rr) -- (rring.west)
    node[midway, below=2pt, lbl] {reader};
  \draw[flow, srcIO] (scratch.south east) to[out=-25, in=-152]
    node[pos=0.48, below=1pt, lbl] {or} ([xshift=-2mm]rring.south);
  \draw[flow, srcIO] (rring.east) -- (kern.west |- 0,\rr)
    node[midway, below=1.5pt, lbl] {compute\\stream};
  \draw[flow, srcIO] ([xshift=-2.5mm]kern.south) -- ([xshift=-2.5mm]dtier.north);
  \draw[flow, srcIO] ([xshift= 2.5mm]dtier.north) -- ([xshift= 2.5mm]kern.south);

  \foreach \y/\t in {\rc/cores, \rw/writes, \rr/reads}
    \node[font=\dscript, anchor=east] at (-0.1,\y) {\t};
\end{tikzpicture}

%% file: figures/tikz/kicked_stack.tex
\begin{tikzpicture}[x=1cm, y=1cm]
  \tikzset{
    oper/.style={tensor, fill=srcState},
    later/.style={tensor, fill=srcLayer!40, draw=srcInk!55},
    zerocap/.style={circle, draw=srcInk, line width=0.6pt, fill=white, minimum size=2.4mm,
      inner sep=0pt},
    gate/.style={rectangle, rounded corners=1pt, draw=srcInk, line width=0.5pt,
      fill=white, inner sep=1pt, font=\dtiny},
    trivial/.style={line width=0.5pt, draw=srcInk!45, dash pattern=on 1.5pt off 1pt},
    farleg/.style={line width=0.6pt, draw=srcInk!55, dash pattern=on 1pt off 1.5pt},
  }
  \def\xs{0/0, 2/0.6, 3/1.2, 4/1.8, 5/3.0}

  \node[panel] at (-1.3,2.62) {(a)};
  \foreach \r/\y/\sty/\lab in {
      t/2.05/zerocap/{\langle0|},
      f/1.55/later/{U_1^\dagger},
      a/0.6/layer/{U_s^\dagger},
      o/0/oper/{O},
      b/-0.6/layer/{U_s},
      g/-1.55/later/{U_1},
      z/-2.05/zerocap/{|0\rangle}} {
    \node[idx, anchor=east] at (-0.35,\y) {$\lab$};
    \foreach \i/\x in \xs \node[\sty] (\r\i) at (\x,\y) {};
  }
  \node[idx, anchor=east] at (-0.42,1.12) {$\vdots$};
  \node[idx, anchor=east] at (-0.42,-0.98) {$\vdots$};
  \foreach \r/\sty in {f/farleg, a/bond, o/bond, b/bond, g/farleg} {
    \foreach \i/\j in {0/2, 2/3, 3/4} \draw[\sty] (\r\i) -- (\r\j);
    \draw[\sty] (\r4) -- ($(\r4)+(0.33,0)$);
    \draw[\sty] ($(\r5)-(0.33,0)$) -- (\r5);
    \node[font=\dscript] at ($(\r4)!0.5!(\r5)$) {$\cdots$};
  }
  \foreach \i/\x in \xs {
    \draw[leg] (t\i) -- (f\i);
    \draw[farleg] (f\i) -- (a\i);
    \draw[leg] (a\i) -- (o\i) -- (b\i);
    \draw[farleg] (b\i) -- (g\i);
    \draw[leg] (g\i) -- (z\i);
  }
  \node[idx, anchor=south] at (0.3,0.62) {$2$};
  \node[idx, anchor=south] at (0.3,0.02) {$\chi$};
  \node[idx, anchor=south] at (0.3,-0.58) {$2$};
  \draw[decorate, decoration={brace, amplitude=4pt}, line width=0.6pt]
    (3.4,0.82) -- (3.4,-0.82);
  \node[idx, anchor=west, align=left] at (3.55,0)
    {stack, $k=3$:\\one SRC sweep\\$\to$ bond $\chi$};
  \draw[op] (3.4,0.95) -- (3.4,1.4);
  \draw[op] (3.4,-0.95) -- (3.4,-1.4);
  \node[idx, anchor=west, text=srcInk!70] at (3.5,1.17) {next: $U_{s-1}^\dagger$};
  \node[idx, anchor=west, text=srcInk!70] at (3.5,-1.17) {next: $U_{s-1}$};

  \begin{scope}[yshift=-3.2cm]
    \node[panel] at (-1.25,0.8) {(b)};
    \node[idx, anchor=east] at (-0.35,0) {$U_s$};
    \foreach \i in {1,...,8} \node[layer] (b\i) at ({(\i-1)*0.6},0) {};
    \foreach \i/\j/\bend in {1/2/75, 3/6/35, 7/8/75}
      \draw[line width=0.6pt, draw=srcInk] (b\i) to[bend left=\bend]
        node[gate, midway] {$ZZ$} (b\j);
    \foreach \i/\j/\d in {1/2/2, 3/4/2, 4/5/2, 5/6/2, 7/8/2}
      \draw[bond] (b\i) -- (b\j) node[idx, midway, below=1pt] {$\d$};
    \foreach \i/\j in {2/3, 6/7}
      \draw[trivial] (b\i) -- (b\j) node[idx, midway, below=1pt, text=srcInk!60] {$1$};
    \node[idx, anchor=west, align=left] at (4.55,0.1)
      {gates do not overlap\\$\Rightarrow$ bond $\le2$};
  \end{scope}
\end{tikzpicture}

%% file: main.bbl
\begin{thebibliography}{36}%
\makeatletter
\providecommand \@ifxundefined [1]{%
 \@ifx{#1\undefined}
}%
\providecommand \@ifnum [1]{%
 \ifnum #1\expandafter \@firstoftwo
 \else \expandafter \@secondoftwo
 \fi
}%
\providecommand \@ifx [1]{%
 \ifx #1\expandafter \@firstoftwo
 \else \expandafter \@secondoftwo
 \fi
}%
\providecommand \natexlab [1]{#1}%
\providecommand \enquote  [1]{``#1''}%
\providecommand \bibnamefont  [1]{#1}%
\providecommand \bibfnamefont [1]{#1}%
\providecommand \citenamefont [1]{#1}%
\providecommand \href@noop [0]{\@secondoftwo}%
\providecommand \href [0]{\begingroup \@sanitize@url \@href}%
\providecommand \@href[1]{\@@startlink{#1}\@@href}%
\providecommand \@@href[1]{\endgroup#1\@@endlink}%
\providecommand \@sanitize@url [0]{\catcode `\\12\catcode `\$12\catcode `\&12\catcode `\#12\catcode `\^12\catcode `\_12\catcode `\%12\relax}%
\providecommand \@@startlink[1]{}%
\providecommand \@@endlink[0]{}%
\providecommand \url  [0]{\begingroup\@sanitize@url \@url }%
\providecommand \@url [1]{\endgroup\@href {#1}{\urlprefix }}%
\providecommand \urlprefix  [0]{URL }%
\providecommand \Eprint [0]{\href }%
\providecommand \doibase [0]{https://doi.org/}%
\providecommand \selectlanguage [0]{\@gobble}%
\providecommand \bibinfo  [0]{\@secondoftwo}%
\providecommand \bibfield  [0]{\@secondoftwo}%
\providecommand \translation [1]{[#1]}%
\providecommand \BibitemOpen [0]{}%
\providecommand \bibitemStop [0]{}%
\providecommand \bibitemNoStop [0]{.\EOS\space}%
\providecommand \EOS [0]{\spacefactor3000\relax}%
\providecommand \BibitemShut  [1]{\csname bibitem#1\endcsname}%
\let\auto@bib@innerbib\@empty
\bibitem [{\citenamefont {Cama{\~n}o}\ \emph {et~al.}(2026)\citenamefont {Cama{\~n}o}, \citenamefont {Epperly},\ and\ \citenamefont {Tropp}}]{camano2026src}%
  \BibitemOpen
  \bibfield  {author} {\bibinfo {author} {\bibfnamefont {C.}~\bibnamefont {Cama{\~n}o}}, \bibinfo {author} {\bibfnamefont {E.~N.}\ \bibnamefont {Epperly}},\ and\ \bibinfo {author} {\bibfnamefont {J.~A.}\ \bibnamefont {Tropp}},\ }\href {https://doi.org/10.22331/q-2026-03-10-2022} {\bibfield  {journal} {\bibinfo  {journal} {Quantum}\ }\textbf {\bibinfo {volume} {10}},\ \bibinfo {pages} {2022} (\bibinfo {year} {2026})},\ \Eprint {https://arxiv.org/abs/2504.06475} {arXiv:2504.06475} \BibitemShut {NoStop}%
\bibitem [{\citenamefont {Al~Daas}\ \emph {et~al.}(2023)\citenamefont {Al~Daas}, \citenamefont {Ballard}, \citenamefont {Cazeaux}, \citenamefont {Hallman}, \citenamefont {Mi{\k{e}}dlar}, \citenamefont {Pasha}, \citenamefont {Reid},\ and\ \citenamefont {Saibaba}}]{aldaas2023randomized}%
  \BibitemOpen
  \bibfield  {author} {\bibinfo {author} {\bibfnamefont {H.}~\bibnamefont {Al~Daas}}, \bibinfo {author} {\bibfnamefont {G.}~\bibnamefont {Ballard}}, \bibinfo {author} {\bibfnamefont {P.}~\bibnamefont {Cazeaux}}, \bibinfo {author} {\bibfnamefont {E.}~\bibnamefont {Hallman}}, \bibinfo {author} {\bibfnamefont {A.}~\bibnamefont {Mi{\k{e}}dlar}}, \bibinfo {author} {\bibfnamefont {M.}~\bibnamefont {Pasha}}, \bibinfo {author} {\bibfnamefont {T.~W.}\ \bibnamefont {Reid}},\ and\ \bibinfo {author} {\bibfnamefont {A.~K.}\ \bibnamefont {Saibaba}},\ }\href {https://doi.org/10.1137/21M1451191} {\bibfield  {journal} {\bibinfo  {journal} {SIAM Journal on Scientific Computing}\ }\textbf {\bibinfo {volume} {45}},\ \bibinfo {pages} {A74} (\bibinfo {year} {2023})}\BibitemShut {NoStop}%
\bibitem [{\citenamefont {Al~Daas}\ \emph {et~al.}(2026)\citenamefont {Al~Daas}, \citenamefont {Ballard}, \citenamefont {Grigori}, \citenamefont {Martinez~Aguilar}, \citenamefont {Saibaba},\ and\ \citenamefont {Verma}}]{aldaas2026khatrirao}%
  \BibitemOpen
  \bibfield  {author} {\bibinfo {author} {\bibfnamefont {H.}~\bibnamefont {Al~Daas}}, \bibinfo {author} {\bibfnamefont {G.}~\bibnamefont {Ballard}}, \bibinfo {author} {\bibfnamefont {L.}~\bibnamefont {Grigori}}, \bibinfo {author} {\bibfnamefont {M.}~\bibnamefont {Martinez~Aguilar}}, \bibinfo {author} {\bibfnamefont {A.~K.}\ \bibnamefont {Saibaba}},\ and\ \bibinfo {author} {\bibfnamefont {B.~D.}\ \bibnamefont {Verma}},\ }\href {https://doi.org/10.1137/25M181708X} {\bibfield  {journal} {\bibinfo  {journal} {SIAM Journal on Scientific Computing}\ }\textbf {\bibinfo {volume} {48}},\ \bibinfo {pages} {A2959} (\bibinfo {year} {2026})}\BibitemShut {NoStop}%
\bibitem [{\citenamefont {Paeckel}\ \emph {et~al.}(2019)\citenamefont {Paeckel}, \citenamefont {K{\"o}hler}, \citenamefont {Swoboda}, \citenamefont {Manmana}, \citenamefont {Schollw{\"o}ck},\ and\ \citenamefont {Hubig}}]{paeckel2019timeevolution}%
  \BibitemOpen
  \bibfield  {author} {\bibinfo {author} {\bibfnamefont {S.}~\bibnamefont {Paeckel}}, \bibinfo {author} {\bibfnamefont {T.}~\bibnamefont {K{\"o}hler}}, \bibinfo {author} {\bibfnamefont {A.}~\bibnamefont {Swoboda}}, \bibinfo {author} {\bibfnamefont {S.~R.}\ \bibnamefont {Manmana}}, \bibinfo {author} {\bibfnamefont {U.}~\bibnamefont {Schollw{\"o}ck}},\ and\ \bibinfo {author} {\bibfnamefont {C.}~\bibnamefont {Hubig}},\ }\href {https://doi.org/10.1016/j.aop.2019.167998} {\bibfield  {journal} {\bibinfo  {journal} {Annals of Physics}\ }\textbf {\bibinfo {volume} {411}},\ \bibinfo {pages} {167998} (\bibinfo {year} {2019})}\BibitemShut {NoStop}%
\bibitem [{\citenamefont {Schollw{\"o}ck}(2011)}]{schollwock2011dmrg}%
  \BibitemOpen
  \bibfield  {author} {\bibinfo {author} {\bibfnamefont {U.}~\bibnamefont {Schollw{\"o}ck}},\ }\href {https://doi.org/10.1016/j.aop.2010.09.012} {\bibfield  {journal} {\bibinfo  {journal} {Annals of Physics}\ }\textbf {\bibinfo {volume} {326}},\ \bibinfo {pages} {96} (\bibinfo {year} {2011})}\BibitemShut {NoStop}%
\bibitem [{\citenamefont {Oseledets}(2011)}]{oseledets2011tt}%
  \BibitemOpen
  \bibfield  {author} {\bibinfo {author} {\bibfnamefont {I.~V.}\ \bibnamefont {Oseledets}},\ }\href {https://doi.org/10.1137/090752286} {\bibfield  {journal} {\bibinfo  {journal} {SIAM Journal on Scientific Computing}\ }\textbf {\bibinfo {volume} {33}},\ \bibinfo {pages} {2295} (\bibinfo {year} {2011})}\BibitemShut {NoStop}%
\bibitem [{\citenamefont {Barron}\ \emph {et~al.}(2026)\citenamefont {Barron}, \citenamefont {Mitchell}, \citenamefont {Tripathi}, \citenamefont {Grieco}, \citenamefont {Rosen}, \citenamefont {Pietracaprina}, \citenamefont {Materia}, \citenamefont {Seif}, \citenamefont {Wanisch}, \citenamefont {Panad{\'e}s-Barrueta} \emph {et~al.}}]{barron2026observable}%
  \BibitemOpen
  \bibfield  {author} {\bibinfo {author} {\bibfnamefont {S.~V.}\ \bibnamefont {Barron}}, \bibinfo {author} {\bibfnamefont {B.}~\bibnamefont {Mitchell}}, \bibinfo {author} {\bibfnamefont {V.}~\bibnamefont {Tripathi}}, \bibinfo {author} {\bibfnamefont {F.}~\bibnamefont {Grieco}}, \bibinfo {author} {\bibfnamefont {I.}~\bibnamefont {Rosen}}, \bibinfo {author} {\bibfnamefont {F.}~\bibnamefont {Pietracaprina}}, \bibinfo {author} {\bibfnamefont {D.}~\bibnamefont {Materia}}, \bibinfo {author} {\bibfnamefont {A.}~\bibnamefont {Seif}}, \bibinfo {author} {\bibfnamefont {D.}~\bibnamefont {Wanisch}}, \bibinfo {author} {\bibfnamefont {R.~L.}\ \bibnamefont {Panad{\'e}s-Barrueta}}, \emph {et~al.},\ }\href {https://doi.org/10.48550/arXiv.2607.25998} {\bibinfo {title} {Observable estimation in the absence of classical verification}} (\bibinfo {year} {2026}),\ \Eprint {https://arxiv.org/abs/2607.25998} {arXiv:2607.25998 [quant-ph]} \BibitemShut {NoStop}%
\bibitem [{\citenamefont {Filippov}\ \emph {et~al.}(2023)\citenamefont {Filippov}, \citenamefont {Leahy}, \citenamefont {Rossi},\ and\ \citenamefont {Garc{\'i}a-P{\'e}rez}}]{filippov2023tem}%
  \BibitemOpen
  \bibfield  {author} {\bibinfo {author} {\bibfnamefont {S.}~\bibnamefont {Filippov}}, \bibinfo {author} {\bibfnamefont {M.}~\bibnamefont {Leahy}}, \bibinfo {author} {\bibfnamefont {M.~A.~C.}\ \bibnamefont {Rossi}},\ and\ \bibinfo {author} {\bibfnamefont {G.}~\bibnamefont {Garc{\'i}a-P{\'e}rez}},\ }\href {https://doi.org/10.48550/arXiv.2307.11740} {\bibinfo {title} {Scalable tensor-network error mitigation for near-term quantum computing}} (\bibinfo {year} {2023}),\ \Eprint {https://arxiv.org/abs/2307.11740} {arXiv:2307.11740 [quant-ph]} \BibitemShut {NoStop}%
\bibitem [{\citenamefont {Mangini}\ \emph {et~al.}(2024)\citenamefont {Mangini}, \citenamefont {Cattaneo}, \citenamefont {Cavalcanti}, \citenamefont {Filippov}, \citenamefont {Rossi},\ and\ \citenamefont {Garc{\'i}a-P{\'e}rez}}]{mangini2024noise}%
  \BibitemOpen
  \bibfield  {author} {\bibinfo {author} {\bibfnamefont {S.}~\bibnamefont {Mangini}}, \bibinfo {author} {\bibfnamefont {M.}~\bibnamefont {Cattaneo}}, \bibinfo {author} {\bibfnamefont {D.}~\bibnamefont {Cavalcanti}}, \bibinfo {author} {\bibfnamefont {S.}~\bibnamefont {Filippov}}, \bibinfo {author} {\bibfnamefont {M.~A.~C.}\ \bibnamefont {Rossi}},\ and\ \bibinfo {author} {\bibfnamefont {G.}~\bibnamefont {Garc{\'i}a-P{\'e}rez}},\ }\href {https://doi.org/10.1103/PhysRevResearch.6.033217} {\bibfield  {journal} {\bibinfo  {journal} {Physical Review Research}\ }\textbf {\bibinfo {volume} {6}},\ \bibinfo {pages} {033217} (\bibinfo {year} {2024})}\BibitemShut {NoStop}%
\bibitem [{\citenamefont {Filippov}\ \emph {et~al.}(2024)\citenamefont {Filippov}, \citenamefont {Maniscalco},\ and\ \citenamefont {Garc{\'i}a-P{\'e}rez}}]{filippov2024scalability}%
  \BibitemOpen
  \bibfield  {author} {\bibinfo {author} {\bibfnamefont {S.~N.}\ \bibnamefont {Filippov}}, \bibinfo {author} {\bibfnamefont {S.}~\bibnamefont {Maniscalco}},\ and\ \bibinfo {author} {\bibfnamefont {G.}~\bibnamefont {Garc{\'i}a-P{\'e}rez}},\ }\href {https://doi.org/10.48550/arXiv.2403.13542} {\bibinfo {title} {Scalability of quantum error mitigation techniques: from utility to advantage}} (\bibinfo {year} {2024}),\ \Eprint {https://arxiv.org/abs/2403.13542} {arXiv:2403.13542 [quant-ph]} \BibitemShut {NoStop}%
\bibitem [{\citenamefont {Stoudenmire}\ and\ \citenamefont {White}(2010)}]{stoudenmire2010zipup}%
  \BibitemOpen
  \bibfield  {author} {\bibinfo {author} {\bibfnamefont {E.~M.}\ \bibnamefont {Stoudenmire}}\ and\ \bibinfo {author} {\bibfnamefont {S.~R.}\ \bibnamefont {White}},\ }\href {https://doi.org/10.1088/1367-2630/12/5/055026} {\bibfield  {journal} {\bibinfo  {journal} {New Journal of Physics}\ }\textbf {\bibinfo {volume} {12}},\ \bibinfo {pages} {055026} (\bibinfo {year} {2010})}\BibitemShut {NoStop}%
\bibitem [{\citenamefont {Kressner}\ \emph {et~al.}(2023)\citenamefont {Kressner}, \citenamefont {Vandereycken},\ and\ \citenamefont {Voorhaar}}]{kressner2023streaming}%
  \BibitemOpen
  \bibfield  {author} {\bibinfo {author} {\bibfnamefont {D.}~\bibnamefont {Kressner}}, \bibinfo {author} {\bibfnamefont {B.}~\bibnamefont {Vandereycken}},\ and\ \bibinfo {author} {\bibfnamefont {R.}~\bibnamefont {Voorhaar}},\ }\href {https://doi.org/10.1137/22M1515045} {\bibfield  {journal} {\bibinfo  {journal} {SIAM Journal on Scientific Computing}\ }\textbf {\bibinfo {volume} {45}},\ \bibinfo {pages} {A2610} (\bibinfo {year} {2023})}\BibitemShut {NoStop}%
\bibitem [{\citenamefont {Ouyang}\ \emph {et~al.}(2026)\citenamefont {Ouyang}, \citenamefont {Chi},\ and\ \citenamefont {Chan}}]{ouyang2026hubbard}%
  \BibitemOpen
  \bibfield  {author} {\bibinfo {author} {\bibfnamefont {X.-Y.}\ \bibnamefont {Ouyang}}, \bibinfo {author} {\bibfnamefont {R.}~\bibnamefont {Chi}},\ and\ \bibinfo {author} {\bibfnamefont {G.~K.-L.}\ \bibnamefont {Chan}},\ }\href {https://doi.org/10.48550/arXiv.2608.13805} {\bibinfo {title} {Fast classical simulation of `{Fast}, accurate, high-resolution simulation of large-scale {Fermi}--{Hubbard} models on a digital quantum processor'}} (\bibinfo {year} {2026}),\ \Eprint {https://arxiv.org/abs/2608.13805} {arXiv:2608.13805 [quant-ph]} \BibitemShut {NoStop}%
\bibitem [{\citenamefont {Foder{\`a}}\ \emph {et~al.}(2026)\citenamefont {Foder{\`a}}, \citenamefont {Ritter}, \citenamefont {Shinaoka},\ and\ \citenamefont {von Delft}}]{fodera2026parallel}%
  \BibitemOpen
  \bibfield  {author} {\bibinfo {author} {\bibfnamefont {S.}~\bibnamefont {Foder{\`a}}}, \bibinfo {author} {\bibfnamefont {M.~K.}\ \bibnamefont {Ritter}}, \bibinfo {author} {\bibfnamefont {H.}~\bibnamefont {Shinaoka}},\ and\ \bibinfo {author} {\bibfnamefont {J.}~\bibnamefont {von Delft}},\ }\href {https://doi.org/10.48550/arXiv.2606.23274} {\bibinfo {title} {Parallelized contraction of tensor trains or matrix product operators}} (\bibinfo {year} {2026}),\ \Eprint {https://arxiv.org/abs/2606.23274} {arXiv:2606.23274 [physics.comp-ph]} \BibitemShut {NoStop}%
\bibitem [{\citenamefont {Halko}\ \emph {et~al.}(2011)\citenamefont {Halko}, \citenamefont {Martinsson},\ and\ \citenamefont {Tropp}}]{halko2011finding}%
  \BibitemOpen
  \bibfield  {author} {\bibinfo {author} {\bibfnamefont {N.}~\bibnamefont {Halko}}, \bibinfo {author} {\bibfnamefont {P.-G.}\ \bibnamefont {Martinsson}},\ and\ \bibinfo {author} {\bibfnamefont {J.~A.}\ \bibnamefont {Tropp}},\ }\href {https://doi.org/10.1137/090771806} {\bibfield  {journal} {\bibinfo  {journal} {SIAM Review}\ }\textbf {\bibinfo {volume} {53}},\ \bibinfo {pages} {217} (\bibinfo {year} {2011})}\BibitemShut {NoStop}%
\bibitem [{\citenamefont {Martinsson}\ and\ \citenamefont {Tropp}(2020)}]{martinsson2020rnla}%
  \BibitemOpen
  \bibfield  {author} {\bibinfo {author} {\bibfnamefont {P.-G.}\ \bibnamefont {Martinsson}}\ and\ \bibinfo {author} {\bibfnamefont {J.~A.}\ \bibnamefont {Tropp}},\ }\href {https://doi.org/10.1017/S0962492920000021} {\bibfield  {journal} {\bibinfo  {journal} {Acta Numerica}\ }\textbf {\bibinfo {volume} {29}},\ \bibinfo {pages} {403} (\bibinfo {year} {2020})}\BibitemShut {NoStop}%
\bibitem [{\citenamefont {Cama{\~n}o}\ \emph {et~al.}(2025)\citenamefont {Cama{\~n}o}, \citenamefont {Epperly}, \citenamefont {Meyer},\ and\ \citenamefont {Tropp}}]{camano2025osi}%
  \BibitemOpen
  \bibfield  {author} {\bibinfo {author} {\bibfnamefont {C.}~\bibnamefont {Cama{\~n}o}}, \bibinfo {author} {\bibfnamefont {E.~N.}\ \bibnamefont {Epperly}}, \bibinfo {author} {\bibfnamefont {R.~A.}\ \bibnamefont {Meyer}},\ and\ \bibinfo {author} {\bibfnamefont {J.~A.}\ \bibnamefont {Tropp}},\ }\href {https://doi.org/10.48550/arXiv.2508.21189} {\bibinfo {title} {Faster linear algebra algorithms with structured random matrices}} (\bibinfo {year} {2025}),\ \Eprint {https://arxiv.org/abs/2508.21189} {arXiv:2508.21189 [cs.DS]} \BibitemShut {NoStop}%
\bibitem [{\citenamefont {Smith}\ and\ \citenamefont {Gray}(2018)}]{smith2018opteinsum}%
  \BibitemOpen
  \bibfield  {author} {\bibinfo {author} {\bibfnamefont {D.~G.~A.}\ \bibnamefont {Smith}}\ and\ \bibinfo {author} {\bibfnamefont {J.}~\bibnamefont {Gray}},\ }\href {https://doi.org/10.21105/joss.00753} {\bibfield  {journal} {\bibinfo  {journal} {Journal of Open Source Software}\ }\textbf {\bibinfo {volume} {3}},\ \bibinfo {pages} {753} (\bibinfo {year} {2018})}\BibitemShut {NoStop}%
\bibitem [{\citenamefont {Fukaya}\ \emph {et~al.}(2014)\citenamefont {Fukaya}, \citenamefont {Nakatsukasa}, \citenamefont {Yanagisawa},\ and\ \citenamefont {Yamamoto}}]{fukaya2014choleskyqr2}%
  \BibitemOpen
  \bibfield  {author} {\bibinfo {author} {\bibfnamefont {T.}~\bibnamefont {Fukaya}}, \bibinfo {author} {\bibfnamefont {Y.}~\bibnamefont {Nakatsukasa}}, \bibinfo {author} {\bibfnamefont {Y.}~\bibnamefont {Yanagisawa}},\ and\ \bibinfo {author} {\bibfnamefont {Y.}~\bibnamefont {Yamamoto}},\ }in\ \href {https://doi.org/10.1109/ScalA.2014.11} {\emph {\bibinfo {booktitle} {Proceedings of the 5th Workshop on Latest Advances in Scalable Algorithms for Large-Scale Systems (ScalA)}}}\ (\bibinfo {year} {2014})\ pp.\ \bibinfo {pages} {31--38}\BibitemShut {NoStop}%
\bibitem [{\citenamefont {Fukaya}\ \emph {et~al.}(2020)\citenamefont {Fukaya}, \citenamefont {Kannan}, \citenamefont {Nakatsukasa}, \citenamefont {Yamamoto},\ and\ \citenamefont {Yanagisawa}}]{fukaya2020shifted}%
  \BibitemOpen
  \bibfield  {author} {\bibinfo {author} {\bibfnamefont {T.}~\bibnamefont {Fukaya}}, \bibinfo {author} {\bibfnamefont {R.}~\bibnamefont {Kannan}}, \bibinfo {author} {\bibfnamefont {Y.}~\bibnamefont {Nakatsukasa}}, \bibinfo {author} {\bibfnamefont {Y.}~\bibnamefont {Yamamoto}},\ and\ \bibinfo {author} {\bibfnamefont {Y.}~\bibnamefont {Yanagisawa}},\ }\href {https://doi.org/10.1137/18M1218212} {\bibfield  {journal} {\bibinfo  {journal} {SIAM Journal on Scientific Computing}\ }\textbf {\bibinfo {volume} {42}},\ \bibinfo {pages} {A477} (\bibinfo {year} {2020})}\BibitemShut {NoStop}%
\bibitem [{\citenamefont {Williams}\ \emph {et~al.}(2009)\citenamefont {Williams}, \citenamefont {Waterman},\ and\ \citenamefont {Patterson}}]{williams2009roofline}%
  \BibitemOpen
  \bibfield  {author} {\bibinfo {author} {\bibfnamefont {S.}~\bibnamefont {Williams}}, \bibinfo {author} {\bibfnamefont {A.}~\bibnamefont {Waterman}},\ and\ \bibinfo {author} {\bibfnamefont {D.}~\bibnamefont {Patterson}},\ }\href {https://doi.org/10.1145/1498765.1498785} {\bibfield  {journal} {\bibinfo  {journal} {Communications of the ACM}\ }\textbf {\bibinfo {volume} {52}},\ \bibinfo {pages} {65} (\bibinfo {year} {2009})}\BibitemShut {NoStop}%
\bibitem [{\citenamefont {Valiant}(1990)}]{valiant1990bsp}%
  \BibitemOpen
  \bibfield  {author} {\bibinfo {author} {\bibfnamefont {L.~G.}\ \bibnamefont {Valiant}},\ }\href {https://doi.org/10.1145/79173.79181} {\bibfield  {journal} {\bibinfo  {journal} {Communications of the ACM}\ }\textbf {\bibinfo {volume} {33}},\ \bibinfo {pages} {103} (\bibinfo {year} {1990})}\BibitemShut {NoStop}%
\bibitem [{\citenamefont {Thakur}\ \emph {et~al.}(2005)\citenamefont {Thakur}, \citenamefont {Rabenseifner},\ and\ \citenamefont {Gropp}}]{thakur2005collectives}%
  \BibitemOpen
  \bibfield  {author} {\bibinfo {author} {\bibfnamefont {R.}~\bibnamefont {Thakur}}, \bibinfo {author} {\bibfnamefont {R.}~\bibnamefont {Rabenseifner}},\ and\ \bibinfo {author} {\bibfnamefont {W.}~\bibnamefont {Gropp}},\ }\href {https://doi.org/10.1177/1094342005051521} {\bibfield  {journal} {\bibinfo  {journal} {The International Journal of High Performance Computing Applications}\ }\textbf {\bibinfo {volume} {19}},\ \bibinfo {pages} {49} (\bibinfo {year} {2005})}\BibitemShut {NoStop}%
\bibitem [{\citenamefont {Hoefler}\ and\ \citenamefont {Belli}(2015)}]{hoefler2015benchmarking}%
  \BibitemOpen
  \bibfield  {author} {\bibinfo {author} {\bibfnamefont {T.}~\bibnamefont {Hoefler}}\ and\ \bibinfo {author} {\bibfnamefont {R.}~\bibnamefont {Belli}},\ }in\ \href {https://doi.org/10.1145/2807591.2807644} {\emph {\bibinfo {booktitle} {Proceedings of the International Conference for High Performance Computing, Networking, Storage and Analysis (SC15)}}}\ (\bibinfo {year} {2015})\ pp.\ \bibinfo {pages} {1--12}\BibitemShut {NoStop}%
\bibitem [{\citenamefont {{CINECA}}(2026)}]{leonardo_specs}%
  \BibitemOpen
  \bibfield  {author} {\bibinfo {author} {\bibnamefont {{CINECA}}},\ }\href@noop {} {\bibinfo {title} {{LEONARDO} user documentation: System architecture and hardware details}},\ \bibinfo {howpublished} {\url{https://docs.hpc.cineca.it/hpc/leonardo.html}} (\bibinfo {year} {2026})\BibitemShut {NoStop}%
\bibitem [{\citenamefont {{J{\"u}lich Supercomputing Centre (JSC)}}(2026)}]{jupiter_specs}%
  \BibitemOpen
  \bibfield  {author} {\bibinfo {author} {\bibnamefont {{J{\"u}lich Supercomputing Centre (JSC)}}},\ }\href@noop {} {\bibinfo {title} {{JUPITER} user documentation: Hardware configuration}},\ \bibinfo {howpublished} {\url{https://apps.fz-juelich.de/jsc/hps/jupiter/configuration.html}} (\bibinfo {year} {2026})\BibitemShut {NoStop}%
\bibitem [{\citenamefont {Harris}\ \emph {et~al.}(2020)\citenamefont {Harris}, \citenamefont {Millman}, \citenamefont {van~der Walt} \emph {et~al.}}]{harris2020numpy}%
  \BibitemOpen
  \bibfield  {author} {\bibinfo {author} {\bibfnamefont {C.~R.}\ \bibnamefont {Harris}}, \bibinfo {author} {\bibfnamefont {K.~J.}\ \bibnamefont {Millman}}, \bibinfo {author} {\bibfnamefont {S.~J.}\ \bibnamefont {van~der Walt}}, \emph {et~al.},\ }\href {https://doi.org/10.1038/s41586-020-2649-2} {\bibfield  {journal} {\bibinfo  {journal} {Nature}\ }\textbf {\bibinfo {volume} {585}},\ \bibinfo {pages} {357} (\bibinfo {year} {2020})}\BibitemShut {NoStop}%
\bibitem [{\citenamefont {Okuta}\ \emph {et~al.}(2017)\citenamefont {Okuta}, \citenamefont {Unno}, \citenamefont {Nishino}, \citenamefont {Hido},\ and\ \citenamefont {Loomis}}]{okuta2017cupy}%
  \BibitemOpen
  \bibfield  {author} {\bibinfo {author} {\bibfnamefont {R.}~\bibnamefont {Okuta}}, \bibinfo {author} {\bibfnamefont {Y.}~\bibnamefont {Unno}}, \bibinfo {author} {\bibfnamefont {D.}~\bibnamefont {Nishino}}, \bibinfo {author} {\bibfnamefont {S.}~\bibnamefont {Hido}},\ and\ \bibinfo {author} {\bibfnamefont {C.}~\bibnamefont {Loomis}},\ }in\ \href {http://learningsys.org/nips17/assets/papers/paper_16.pdf} {\emph {\bibinfo {booktitle} {Proceedings of the Workshop on Machine Learning Systems (LearningSys) at the 31st Conference on Neural Information Processing Systems}}}\ (\bibinfo {year} {2017})\BibitemShut {NoStop}%
\bibitem [{\citenamefont {Gray}(2018)}]{gray2018quimb}%
  \BibitemOpen
  \bibfield  {author} {\bibinfo {author} {\bibfnamefont {J.}~\bibnamefont {Gray}},\ }\href {https://doi.org/10.21105/joss.00819} {\bibfield  {journal} {\bibinfo  {journal} {Journal of Open Source Software}\ }\textbf {\bibinfo {volume} {3}},\ \bibinfo {pages} {819} (\bibinfo {year} {2018})}\BibitemShut {NoStop}%
\bibitem [{\citenamefont {Ba{\~n}uls}\ \emph {et~al.}(2011)\citenamefont {Ba{\~n}uls}, \citenamefont {Cirac},\ and\ \citenamefont {Hastings}}]{banuls2011thermalization}%
  \BibitemOpen
  \bibfield  {author} {\bibinfo {author} {\bibfnamefont {M.~C.}\ \bibnamefont {Ba{\~n}uls}}, \bibinfo {author} {\bibfnamefont {J.~I.}\ \bibnamefont {Cirac}},\ and\ \bibinfo {author} {\bibfnamefont {M.~B.}\ \bibnamefont {Hastings}},\ }\href {https://doi.org/10.1103/PhysRevLett.106.050405} {\bibfield  {journal} {\bibinfo  {journal} {Physical Review Letters}\ }\textbf {\bibinfo {volume} {106}},\ \bibinfo {pages} {050405} (\bibinfo {year} {2011})}\BibitemShut {NoStop}%
\bibitem [{\citenamefont {Kim}\ \emph {et~al.}(2023)\citenamefont {Kim}, \citenamefont {Eddins}, \citenamefont {Anand}, \citenamefont {Wei}, \citenamefont {van~den Berg}, \citenamefont {Rosenblatt}, \citenamefont {Nayfeh}, \citenamefont {Wu}, \citenamefont {Zaletel}, \citenamefont {Temme},\ and\ \citenamefont {Kandala}}]{kim2023utility}%
  \BibitemOpen
  \bibfield  {author} {\bibinfo {author} {\bibfnamefont {Y.}~\bibnamefont {Kim}}, \bibinfo {author} {\bibfnamefont {A.}~\bibnamefont {Eddins}}, \bibinfo {author} {\bibfnamefont {S.}~\bibnamefont {Anand}}, \bibinfo {author} {\bibfnamefont {K.~X.}\ \bibnamefont {Wei}}, \bibinfo {author} {\bibfnamefont {E.}~\bibnamefont {van~den Berg}}, \bibinfo {author} {\bibfnamefont {S.}~\bibnamefont {Rosenblatt}}, \bibinfo {author} {\bibfnamefont {H.}~\bibnamefont {Nayfeh}}, \bibinfo {author} {\bibfnamefont {Y.}~\bibnamefont {Wu}}, \bibinfo {author} {\bibfnamefont {M.}~\bibnamefont {Zaletel}}, \bibinfo {author} {\bibfnamefont {K.}~\bibnamefont {Temme}},\ and\ \bibinfo {author} {\bibfnamefont {A.}~\bibnamefont {Kandala}},\ }\href {https://doi.org/10.1038/s41586-023-06096-3} {\bibfield  {journal} {\bibinfo  {journal} {Nature}\ }\textbf {\bibinfo {volume} {618}},\ \bibinfo {pages} {500} (\bibinfo {year} {2023})}\BibitemShut {NoStop}%
\bibitem [{\citenamefont {Tindall}\ \emph {et~al.}(2024)\citenamefont {Tindall}, \citenamefont {Fishman}, \citenamefont {Stoudenmire},\ and\ \citenamefont {Sels}}]{tindall2024eagle}%
  \BibitemOpen
  \bibfield  {author} {\bibinfo {author} {\bibfnamefont {J.}~\bibnamefont {Tindall}}, \bibinfo {author} {\bibfnamefont {M.}~\bibnamefont {Fishman}}, \bibinfo {author} {\bibfnamefont {E.~M.}\ \bibnamefont {Stoudenmire}},\ and\ \bibinfo {author} {\bibfnamefont {D.}~\bibnamefont {Sels}},\ }\href {https://doi.org/10.1103/PRXQuantum.5.010308} {\bibfield  {journal} {\bibinfo  {journal} {PRX Quantum}\ }\textbf {\bibinfo {volume} {5}},\ \bibinfo {pages} {010308} (\bibinfo {year} {2024})}\BibitemShut {NoStop}%
\bibitem [{\citenamefont {Begu{\v{s}}i{\'c}}\ \emph {et~al.}(2024)\citenamefont {Begu{\v{s}}i{\'c}}, \citenamefont {Gray},\ and\ \citenamefont {Chan}}]{begusic2024converged}%
  \BibitemOpen
  \bibfield  {author} {\bibinfo {author} {\bibfnamefont {T.}~\bibnamefont {Begu{\v{s}}i{\'c}}}, \bibinfo {author} {\bibfnamefont {J.}~\bibnamefont {Gray}},\ and\ \bibinfo {author} {\bibfnamefont {G.~K.-L.}\ \bibnamefont {Chan}},\ }\href {https://doi.org/10.1126/sciadv.adk4321} {\bibfield  {journal} {\bibinfo  {journal} {Science Advances}\ }\textbf {\bibinfo {volume} {10}},\ \bibinfo {pages} {eadk4321} (\bibinfo {year} {2024})}\BibitemShut {NoStop}%
\bibitem [{\citenamefont {Anand}\ \emph {et~al.}(2023)\citenamefont {Anand}, \citenamefont {Temme}, \citenamefont {Kandala},\ and\ \citenamefont {Zaletel}}]{anand2023zne}%
  \BibitemOpen
  \bibfield  {author} {\bibinfo {author} {\bibfnamefont {S.}~\bibnamefont {Anand}}, \bibinfo {author} {\bibfnamefont {K.}~\bibnamefont {Temme}}, \bibinfo {author} {\bibfnamefont {A.}~\bibnamefont {Kandala}},\ and\ \bibinfo {author} {\bibfnamefont {M.}~\bibnamefont {Zaletel}},\ }\href {https://doi.org/10.48550/arXiv.2306.17839} {\bibinfo {title} {Classical benchmarking of zero noise extrapolation beyond the exactly-verifiable regime}} (\bibinfo {year} {2023}),\ \Eprint {https://arxiv.org/abs/2306.17839} {arXiv:2306.17839 [quant-ph]} \BibitemShut {NoStop}%
\bibitem [{\citenamefont {Gidney}(2021)}]{gidney2021stim}%
  \BibitemOpen
  \bibfield  {author} {\bibinfo {author} {\bibfnamefont {C.}~\bibnamefont {Gidney}},\ }\href {https://doi.org/10.22331/q-2021-07-06-497} {\bibfield  {journal} {\bibinfo  {journal} {Quantum}\ }\textbf {\bibinfo {volume} {5}},\ \bibinfo {pages} {497} (\bibinfo {year} {2021})}\BibitemShut {NoStop}%
\bibitem [{\citenamefont {{Algorithmiq}}(2026)}]{portc}%
  \BibitemOpen
  \bibfield  {author} {\bibinfo {author} {\bibnamefont {{Algorithmiq}}},\ }\href@noop {} {\bibinfo {title} {portc: Parallel out-of-core randomized tensor-network compression}},\ \bibinfo {howpublished} {\url{https://github.com/Algorithmiq/portc}} (\bibinfo {year} {2026}),\ \bibinfo {note} {{Apache}-2.0 licence}\BibitemShut {NoStop}%
\end{thebibliography}%
